\documentclass[preprint,eqsecnum,preprintnumbers,nofootinbib,byrevtex,prd,aps,showpacs,showkeys,groupedaddress,floatfix]{revtex4}
\usepackage{bm}
\usepackage{graphics}
\usepackage{graphicx}
\usepackage{epsfig}
\usepackage{amssymb}
\usepackage{amsmath}
\begin{document}

%\large
\title{Higher Twist Effects In Proton-Proton Collisions}
\author{A.~I.~Ahmadov$^{1,2}$} \email{E-mail:ahmadovazar@yahoo.com}
\author{I.~Boztosun$^{1}$}
\author{R.~Kh.~Muradov$^{2}$}%
\author{A.~Soylu$^{1}$}
\author{E.~A.~Dadashov$^{2}$}%
\affiliation{$^{1}$Department of Physics, Faculty of Arts and Sciences,
Erciyes University, Kayseri, Turkey \\ $^{2}$ Baku State University,
Z. Khalilov st. 23, AZ-1148, Baku, Azerbaijan}
\date{\today}
\begin{abstract} In this article, we investigate the contribution of
the high twist Feynman diagrams to the large-$p_T$ pion production
cross section in proton-proton collisions and we present the general
formulae for the high and leading twist differential cross sections.
The pion wave function where two non-trivial Gegenbauer coefficients
$a_2$ and $a_4$ have been extracted from the CLEO data, two other
pion model wave functions, $P_2$, $P_3$, the asymptotic and the
Chernyak-Zhitnitsky wave functions are used in the calculations. The
results of all the calculations reveal that the high twist cross
sections, the ratios $R$, $r$, the dependence transverse momentum
$p_T$ and the rapidity $y$ of pion in the $\Phi_{CLEO }(x,Q^2)$ wave
function case is very close to the  $\Phi_{asy}(x)$ asymptotic wave
function case. It is shown that the high twist contribution to the
cross section depends on the choice of the meson wave functions.
\end{abstract}
\pacs{12.38.-t, 13.60.Le, 14.40.Aq, 13.87.Fh, } \keywords{leading
twist, high twist, pion wave function} \maketitle

\section{\bf Introduction}

During the last few years, a great deal of progress has been made in
the investigation of the properties of hadronic wave
functions[1-12]. The pion wave functions (also called distribution
amplitudes -DA) [1] play a key role in the hard-scattering QCD
processes because they encapsulate the essential nonperturbative
features of the pion's internal structure in terms of the parton's
longitudinal momentum fractions $x_i$. Meson wave functions have
been extensively studied by using QCD sum rules. The original
suggestion by Chernyak and Zhitnitsky of a "double-humped" wave
function of the pion at a low scale, far from the asymptotic form,
was based on an extraction of the first few moments from a standard
QCD sum rule approach[5], which has been criticized and revised in
Refs.[6,7]. Subsequently, a number of authors have proposed and
studied the modified versions of meson [7,8] and baryon wave
functions [9,10]. Interesting nucleon functions have been
constructed in the context of the quark-diquark model[11].
Additional arguments in favour of a form of the pion wave functions
close to the asymptotic one have come from the analysis of the
transition form factor $\gamma\gamma^{\star}\to\pi^0$ [12]. The
measurements of this form factor by the CLEO collaboration are
consistent with a near-asymptotic form of the wave function[13]. In
[14], the leading-twist wave function of the pion at a low
normalization point is calculated in the effective low-energy theory
derived from the instanton vacuum. These results for the pion wave
function at the low normalization point are close to the asymptotic
form and consistent with the CLEO measurements. The  authors have
obtained a shape substantially different from the
Chernyak-Zhitnitsky one because they have chosen a significantly
smaller value of the second moment, and, more importantly, they have
taken  all the moments of the wave function into account. Their
results support the conclusions reached previously in Refs.[6,7].
The QCD factorization theorems predict that the hadron-hadron cross
section can be obtained by the convolution of parton distribution
functions and a cross section of the corresponding hard scattering
subprocess. The parton distributions are nonperturbative,
process-independent quantities, which are specific to any given
hadron. The hard scattering cross sections are independent of all
long distance effects and can be found by means of pQCD. In the
framework of pQCD, the higher order corrections to the hard
scattering, and therefore to the hadron-hadron process cross
sections, have been calculated [15]. These corrections are large and
change the leading order results considerably. Other corrections to
the hadron-hadron process cross sections and its different
characteristics come from the higher twist (HT) terms. Thus,
explicit HT effects associated with a meson bound state in the
process $\pi N\to\gamma^{\star}X$ and $\pi N\to\gamma X$  and their
influence on the decay angular distribution have been found by
Brodsky and Berger in Ref.[16]. It is important to note that the
term "twist"("twist"- means dimension minus Lorentz spin) is one of
the characteristics of composite operators that occur in operator
product expansion (OPE) in deep inelastic scattering (DIS). In DIS,
the higher twist (twist$>$2) terms in OPE are associated with power
suppressed corrections. On the other hand, in the various hard
hadron-hadron scatterings- \emph{i.e.} in the Drell-Yan process
lacking an (OPE) description- the higher twist terms refer to
contributions which are suppressed as $O(1/Q^2)$, relative to the
scale of the hard scatttering. Therefore, in order to calculate the
power suppressed corrections to the hadron-hadron collisions cross
sections, the Feynman diagram approach should be used. Indeed, in
the context of this method, the factorization of the HT
contributions of $O(1/Q^2)$ in hadron collisions has been proven
[15]. It is well known that in hadron-hadron  scattering at the
$O(1/Q^4)$ level, the factorization fails, which has been
demonstrated by the existence of non-cancelling infrared divergences
at two loops and by finite terms at one loop[17]. So, the extension
of the factorization to $O(1/Q^2)$ corrections in a large class of
hadronic processes helps to place their existing treatments [16,18]
on a solid foundation. In [15], the leading $1/Q^2$ corrections to
the Drel-Yan cross-section have also been obtained. It is worth
noting that the normalization of the $O(1/Q^2)$ longitudinal
structure functions in the Drell-Yan cross-section are determined by
higher twist longitudinal structure functions in DIS. In other
words, the $1/Q^2$ corrections to the Drell-Yan process can be
expressed in terms of the same multiparton correlations as in the
DIS one.

On the other hand, the results obtained are consistent with the
lowest order calculation of the pion-hadron scattering carried out
by Brodsky and Berger in Ref.[16]. By taking these points into
account, it may be asserted that the analysis of the higher twist
effects  on the dependence of the pion wave function  in pion
production at proton-proton collisions are significant in both
theoretical and experimental studies.

Another important aspect of this study is the choice of the meson
model wave functions. In this respect, the contribution of the high
twist Feynman diagrams to a pion production cross section in
proton-proton collisions has been computed by using various pion
wave functions. Also, the leading and high twist contributions have
been estimated and compared to each other. Within this context, this
paper is organized as follows: in section \ref{ht}, we provide some
formulae for the calculation of the contribution of the high twist
diagrams. In section  \ref{lt}, we provide the formulae for the
calculation of the contribution of the leading twist diagrams and in
section  \ref{results}, we present the numerical results for the
cross section and discuss the dependence of the cross section on the
pion wave functions. We state our conclusions in section
\ref{conc}.

\section{CONTRIBUTION OF THE HIGH TWIST DIAGRAMS}
\label{ht} The high twist Feynman diagrams, which describe the
subprocess $q_1+\bar{q}_{2} \to \pi^{+}(\pi^{-})+\gamma$ for the
pion production in the proton-proton collision are shown in Fig.1.
In the high twist diagrams, the pion of a proton quark is  directly
observed. Their $1/Q^2$ power suppression is caused by a hard gluon
exchange between pion constituents. The amplitude for this
subprocess can be found by means of the Brodsky-Lepage formula [19]

\begin{equation}
M(\hat s,\hat
t)=\int_{0}^{1}{dx_1}\int_{0}^{1}dx_2\delta(1-x_1-x_2)\Phi_{\pi}(x_1,x_2,Q^2)T_{H}(\hat
s,\hat t;x_1,x_2).
\end{equation}

In Eq.(2.1), $T_H$ is  the sum of the graphs contributing to the
hard-scattering part of the subprocess. The hard-scattering part for
the subprocess under consideration is $q_1+\bar{q}_{2} \to
(q_{1}\bar{q}_2)+\gamma$, where a quark and antiquark form a
pseudoscalar, color-singlet state $(q_1\bar{q}_2)$. The
hard-scattering amplitude $T_{H}(\hat s,\hat t;x_1,x_2)$ depends on
a process and can be obtained in the framework of pQCD, whereas the
wave function $\Phi_{\pi}(x_1,x_2,Q^2)$ describes all the
non-perturbative and process-independent effects of hadronic
binding. The hadron wave function gives the amplitude for finding
partons (quarks, gluons) carrying the longitudinal fractional
momenta $\textbf{x}=(x_1,x_2,....x_n)$ and virtualness up to $Q^2$
within the hadron and, in general, includes all Fock states with
quantum numbers of the hadron. But only the lowest Fock state
($q_1\bar{q}_{2}$-for mesons, $uud$-for proton, \emph{etc.})
contributes to the leading scaling behavior, other Fock state
contributions are suppressed by powers of $1/Q^2$. In our work, we
have restricted ourselves to considering the lowest Fock state for a
meson. Then $\textbf{x}= x_1,x_2$ and $x_1+x_2=1$. This approach can
be applied not only to the investigation of exclusive processes, but
also to the calculation of higher twist corrections to some
inclusive processes. The $q_{1}\overline{q}_{2}$ spin state used in
computing $T_H$ may be written in the form

\begin{equation}
\sum_{s_{1},s_{2}}
\frac{u_{s_1}({x}_{1}p_{M})\overline{v}_{s_{2}}({x}_{2}p_{M})}{\sqrt{x_1}
\sqrt{x_2}}\cdot N_{s_{1}s_{2}}^s=\left\{\begin{array}{ccc}
\frac{{\gamma}_{5}\hat {p}_{\pi}}{\sqrt{2}},\,\,\pi,\\\frac{\hat
{p}_{M}}{\sqrt{2}},\,\,\rho_L\,\,helicity \, 0,\\
\mp\frac{{\varepsilon}_{\mp}\hat {p}_{M}}{\sqrt{2}},\,\,
\rho{_T}\,\,helicity \pm1,\end{array}\right.
\end{equation}

where $\varepsilon_{\pm}=\mp(1/\sqrt{2})(0,1,\pm i,0)$ in a frame
with $(p_M)_{1,2}=0$ and the $N_{s_{1}s_{2}}^s$ project out a state
of spins $s$, and $p_{M}$ is the four-momentum of the final meson.
In our calculation, we have neglected the pion and the proton
masses. Turning to extracting the contributions of the high twist
subprocesses, there are many kinds of leading twist subprocesses in
$pp$ collisions as the background of the high twist subprocess
$q_1+q_2 \to \pi^{+}(or\,\, \pi^-)+\gamma$, such as $q+\bar{q} \to
\gamma+g(g \to \pi^{+}(\pi^{-}))$, $q+g \to \gamma+q(q \to
\pi^{+}(\pi^{-}))$, $\bar{q}+g \to \gamma+\bar{q}g(\bar{q} \to
\pi^{+}(\pi^{-}))$ and so on. The contributions from these leading
twist subprocesses strongly depend on some phenomenological factors,
for example, quark and gluon distribution functions in proton and
fragmentation functions of various constituents \emph{etc}. Most of
these factors have not been well determined, neither theoretically
nor experimentally. Thus they cause very large uncertainty in the
computation of the cross section of process $pp \to \pi^{+}(or\,\,
\pi^{-})+\gamma +X$. In general, the magnitude of this uncertainty
is much larger than the sum of all the high twist contributions, so
it is very difficult to extract the high twist contributions.

The Mandelstam invariant variables for subprocesses $q_1+\bar{q}_{2} \to \pi^{+}(\pi^{-})+\gamma$ are defined as
\begin{equation}
\hat s=(p_1+p_2)^2,\quad \hat t=(p_1-p_{\pi})^2,\quad \hat
u=(p_1-p_{\gamma})^2.
\end{equation}

In our calculation, we have also neglected the quark masses. We have
aimed to calculate the pion production cross section and to fix the
differences due to the use of various pion model functions. We have
used five  different functions: the asymptotic wave function ASY,
the Chernyak-Zhitnitsky [2,5], $P_2, P_3$ model functions [7,8] and
the wave function in which two non-trivial Gegenbauer coefficients
$a_2$ and $a_4$ have been extracted from the CLEO data on the
$\gamma\gamma^{\star} \to \pi^0$ transition form factor [20]. In
ref.[20], the authors have used the QCD light-cone sum rules
approach and have included into their analysis the NLO perturbative
and twist-four corrections. They found that in the model with two
nonasymptotic terms, at the scale $\mu_0=2.4 GeV$.

$$
\Phi_{asy}(x)=\sqrt{3}f_{\pi}x(1-x),\quad
\Phi_{CZ}(x,\mu_{0}^2)=5\Phi_{asy}(2x-1)^2,\\
$$
$$
\Phi_{P_2}(x,\mu_{0}^2)=\Phi_{asy}(x)[-0.1821+5.91(2x-1)^2],\\
$$
$$
\Phi_{P_3}(x,\mu_{0}^2)=\Phi_{asy}(x)[0.6016-4.659(2x-1)^2+15.52(2x-1)^4],
$$
\begin{equation}
\Phi_{CLEO}(x,\mu_{0}^2)=\Phi_{asy}(x)[1+0.19C_{2}^{3/2}(2x-1)-0.14C_{4}^{3/2}(2x-1)],
\end{equation}

where $f_{\pi}$=0.0923 GeV is the pion decay constant. Here, we have
denoted by $x\equiv x_1$, the longitudinal fractional momentum
carried by the quark within the meson. Then, $x_2=1-x$ and
$x_1-x_2=2x-1$. The pion wave function is symmetric under
replacement $x_1-x_2\leftrightarrow x_2-x_1$. The values of the pion
wave function moments $<\xi^n>$ are defined as
\begin{equation}
<\xi^n>=\int_{-1}^{1}d\xi \xi^n\widetilde{\Phi}_{\pi}(\xi)
\end{equation}
Here, $\widetilde{\Phi}_{\pi}(\xi)$ is the model function without
$f_{\pi}$ and  $\xi=x_1-x_2$. The pion wave function moments have
been calculated by means of the QCD sum rules method by Chernyak and
Zhitnitsky at the normalization point $\mu_{0}=0.5GeV$. They are
equal to
\begin{equation}
<\xi^0>_{{\mu}_{0}}=1,\,\,\,<\xi^2>_{{\mu}_{0}}=0.44,\,\,<\xi^4>_{{\mu}_{0}}=0.27
\end{equation} The Chernyak-Zhitnitsky pion model wave function has
the following moments
\begin{equation}
<\xi^0>_{{\mu}_{0}}=1,\,\,\,<\xi^2>_{{\mu}_{0}}=0.43,\,\,<\xi^4>_{{\mu}_{0}}=0.24
\end{equation}
It is interesting to note that the corresponding moments of the
asymptotic wave function differ considerably from those in
Eqs.(2.6), (2.7)
\begin{equation}
<\xi^0>_{{\mu}_{0}}=1,\,\,\,<\xi^2>_{{\mu}_{0}}=0.20,\,\,<\xi^4>_{{\mu}_{0}}=0.086
\end{equation}
 This means that the realistic pion wave function is much wider than the asymptotic one
[5,21]. We have also used two other functions $P_2$ and $P_3$. The
wave function $P_2$ is a quadratic polynomial. Its free parameter,
which is defined as the ratio of the coefficients of the constant
and quadratic terms, is fixed to give the best fit to the values of
the 2nd and 4th moments of the wave function. The model function
$P_2$ has the same shape as the $CZ$ one, the difference being in
the constant term. The function $P_3$ is a polynomial in which the
number of free parameters is equal to the number of independent
moments. These parameters are completely fixed by the same moment of
the pion wave function. $P_3$ has a form different from that of the
$CZ$ and $P_2$ ones. All these wave functions, of course, have the
same zeroth moment by construction. Thus, the comparison of the
predictions obtained by using the $CZ$ and $P_2$ ones with the ones
obtained by means of $P_3$ enables us to determine the sensitivity
of the predictions to the form of the wave function.

 The model functions can be written as
$$
\Phi_{asy}(x)=\sqrt{3}f_{\pi}x(1-x),
$$
$$
\Phi_{CZ}(x,\mu_{0}^2)=\Phi_{asy}(x)\left[C_{0}^{3/2}(2x-1)+\frac{2}{3}C_{2}^{3/2}(2x-1)\right],
$$
\begin{equation}
\Phi_{P_2}(x,\mu_{0}^2)=\Phi_{asy}(x)\left[C_{0}^{3/2}(2x-1)+0.788C_{2}^{3/2}(2x-1)\right],\\
\end{equation}
$$
\Phi_{P_3}(x,\mu_{0}^2)=\Phi_{asy}(x)\left[C_{0}^{3/2}(2x-1)+0.7584C_{2}^{3/2}(2x-1)+
0.3942C_{4}^{3/2}(2x-1)\right],\\
$$
$$
\Phi_{CLEO}(x,\mu_{0}^2)=\Phi_{asy}(x)\left[C_{0}^{3/2}(2x-1)+0.19C_{2}^{3/2}(2x-1)-0.14C_{4}^{3/2}(2x-1)\right],
$$
$$
C_{0}^{3/2}(2x-1)=1,\,\,C_{2}^{3/2}(2x-1)=\frac{3}{2}(5(2x-1)^2-1),
$$
$$
C_{4}^{3/2}(2x-1)=\frac{15}{8}(21(2x-1)^4-14(2x-1)^2+1).
$$
It may be seen that the pion wave function extracted from the
experimental data depends on the methods used and their accuracy.
Although one may claim that the meson wave function is a
process-independent quantity, describing the internal structure of
the meson itself, the exploration of different exclusive processes
with the same meson leads to a variety of wave functions. This means
that the methods employed have shortcomings or do not encompass all
the mechanisms important for a given process. Such a situation is
pronounced in the case of the pion. It is known that the pion wave
function (distribution amplitude-DA) can be expanded over the
eigenfunctions of the one-loop Brodsky-Lepage equation, \emph{i.e.},
in terms of the Gegenbauer polynomials $\{C_{n}^{3/2}(2x-1)\},$

\begin{equation}
\Phi_{\pi}(x,Q^2)=\Phi_{asy}(x)\left[1+\sum_{n=2,4..}^{\infty}a_{n}(Q^2)C_{n}^{3/2}(2x-1)\right],
\end{equation}

The evolution of the wave function (DA) on the factorization scale
$Q^2$ is governed by the functions $a_n(Q^2)$,
\begin {equation}
a_n(Q^2)=a_n(\mu_{0}^2)\left[\frac{\alpha_{s}(Q^2)}{\alpha_{s}(\mu_{0}^2)}\right]^{\gamma_n/\beta_0},
\end{equation}
$$
\frac{\gamma_2}{\beta_{0}}=\frac{50}{81},\,\,\,\frac{\gamma_4}{\beta_{0}}=\frac{364}{405},\,\,
n_f=3.
$$

 In Eq.(2.11), $\{\gamma_n\}$ are anomalous dimensions defined by
the expression,

\begin{equation}
\gamma_n=C_F\left[1-\frac{2}{(n+1)(n+2)}+4\sum_{j=2}^{n+1}
\frac{1}{j}\right].
\end{equation}
The constants $a_n(\mu_{0}^2)=a_{n}^0$ are input parameters that
form the shape of the wave functions and which can be extracted from
experimental data or obtained from the nonperturbative QCD
computations at the normalization point $\mu_{0}^2$. The QCD
coupling constant $\alpha_{s}(Q^2)$ at the two-loop approximation is
given by the expression

\begin{equation}
\alpha_{s}(Q^2)=\frac{4\pi}{\beta_0
ln(Q^2/\Lambda^2)}\left[1-\frac{2\beta_1}{\beta_{0}^2}\frac{lnln(Q^2/\Lambda^2)}{ln(Q^2/\Lambda^2)}\right].
\end{equation}
Here, $\Lambda$ is the QCD scale parameter, $\beta_0$ and $\beta_1$
are the QCD beta function one- and two-loop coefficients,
respectively,
$$
\beta_0=11-\frac{2}{3}n_f,\,\,\,\beta_1=51-\frac{19}{3}n_f.
$$
The cross section for the high twist subprocess is given by the
expression
\begin{equation}
\frac{d\sigma}{d\hat t}(\hat s,\hat t,\hat u)=\frac
{8\pi^2\alpha_{E} C_F}{27}\frac{\left[D(\hat t,\hat
u)\right]^2}{{\hat s}^3}\left[\frac{1}{{\hat u}^2}+\frac{1}{{\hat
t}^2}\right],
\end{equation}
where
\begin{equation}
D(\hat t,\hat u)=e_1\hat
t\int_{0}^{1}dx_1\left[\frac{\alpha_{s}(Q_1^2)\Phi_{\pi}(x_1,Q_1^2)}{1-x_1}\right]+e_2\hat
u\int_{0}^{1}dx_1\left[\frac{\alpha_{s}(Q_2^2)\Phi_{\pi}(x_1,Q_2^2)}{1-x_1}\right],
\end{equation}
where $Q_{1}^2=(x_1-1)\hat u,\,\,\,\,Q_{2}^2=-x_1\hat t$,\,\,
represents the momentum squared carried by the hard gluon in Fig.1,
$e_1(e_2)$ is the charge of $q_1(\overline{q}_2)$ and
$C_F=\frac{4}{3}$. The high twist contribution to the large-$p_{T}$
pion production cross section in the process
$pp\to\pi^{+}(\pi^{-})+\gamma$ is [22-24]
\begin{equation}
\Sigma_{M}^{HT}\equiv E\frac{d\sigma}{d^3p}=\int_{0}^{1}\int_{0}^{1}
dx_1 dx_2 G_{{q_{1}}/{h_{1}}}(x_{1})
G_{{q_{2}}/{h_{2}}}(x_{2})\frac{\hat s}{\pi} \frac{d\sigma}{d\hat
t}(q\overline{q}\to \pi\gamma)\delta(\hat s+\hat t+\hat u).
\end{equation}
$$\pi E\frac{d\sigma}{d^3p}=\frac{d\sigma}{dydp_{T}^2},$$
$$\hat s=x_1x_2s,$$
$$\hat t=x_1t,$$
\begin{equation}
\hat u=x_2u,
\end{equation}
$$t= -m_T \sqrt{s} e^{-y}=-p_T \sqrt{s}e^{-y},$$
$$u= -m_T \sqrt {s} e^y=-p_T \sqrt{s}e^{y},$$
$$
x_1=-\frac{x_{2}u}{x_{2}s+t}=\frac{x_{2}p_{T}\sqrt s
e^{y}}{x_{2}s-p_{T}\sqrt s e^{-y}},
$$
$$
x_2=-\frac{x_{1}t}{x_{1}s+u}=\frac{x_{1}p_{T}\sqrt s
e^{-y}}{x_{1}s-p_{T}\sqrt s e^{y}},
$$

 where $m_T$ -- is the transverse mass of pion, which is given by
$$m_T^2=m^2+p_T^2.$$

For a full discussion,  we consider a difference $\Delta^{HT}$
between the high twist cross section combinations
$\Sigma_{\pi^{+}}^{HT}$ and $\Sigma_{\pi^{-}}^{HT}$
\begin{equation}
\Delta_{\pi}^{HT}=\Sigma_{\pi^{+}}^{HT}
-\Sigma_{\pi^{-}}^{HT}=E_{{\pi}^{+}}\frac{d\sigma}{d^3p}
(pp\to\pi^{+}\gamma)-E_{{\pi}^{-}}\frac{d\sigma}{d^3p}(pp \to
\pi^{-}\gamma).
\end{equation}

We have extracted the following high twist subprocesses contributing
to the two covariant cross sections in Eq.(2.16)
\begin{equation}
\frac{{d\sigma}^1}{d\hat t}(u\bar{d} \to \pi^{+}\gamma),\,\,\,
\frac{{d\sigma}^2}{d\hat t}(\bar{d}u \to \pi^{+}\gamma),\,\,\,
\frac{{d\sigma}^3}{d\hat t}(\bar{u}d \to \pi^{-}\gamma),\,\,\,
\frac{{d\sigma}^4}{d\hat t}(d\bar{u} \to \pi^{-}\gamma),\,\,\,
\end{equation}
By charge conjugation invariance, we have
\begin {equation}
\frac{{d\sigma}^1}{d\hat t}(u\bar{d} \to
\pi^{+}\gamma)=\frac{{d\sigma}^3}{d\hat t}(\bar{u}d \to
\pi^{-}\gamma),\,\,and\,\,\, \frac{{d\sigma}^2}{d\hat t}(\bar{d}u
\to \pi^{+}\gamma)= \frac{{d\sigma}^4}{d\hat t}(d\bar{u} \to
\pi^{-}\gamma).
\end{equation}

\section{CONTRIBUTION OF THE  LEADING TWIST DIAGRAMS}\label{lt}
Regarding the high twist corrections to the pion production cross
section, a comparison of our results with leading twist
contributions is crucial. The leading twist subprocesses for the
pion production are quark-antiquark annihilation $q\bar{q} \to
g\gamma$, $g \to \pi^{+}(\pi^{-})$, shown in Fig.2. The
corresponding cross section is easily verified as[22]
\begin{equation}
\frac{d\sigma}{d\hat t}(q\bar{q} \to gq)=\frac{8}{9}\pi\alpha_E
\alpha_s(Q^2)\frac{e_{q}^2}{{\hat s}^2}\left(\frac{\hat t}{\hat
u}+\frac{\hat u}{\hat t}\right),
\end{equation}
For the leading-twist contribution, we find
\begin{equation}
\Sigma_{M}^{LT}\equiv E\frac{d\sigma}{d^3p}=\sum_{q}\int_{0}^{1}
dx_1 dx_2dz G_{{q_{1}}/{h_{1}}}(x_{1})
G_{{q_{2}}/{h_{2}}}(x_{2})D_{g}^{\pi}(z)\frac{\hat s}{\pi
z^2}\frac{d\sigma}{d\hat t}(q\bar{q}\to g\gamma)\delta(\hat s+\hat
t+\hat u),
\end{equation}
where
\begin{equation}
\hat s=x_{1}x_{2}s,\,\,\hat t=\frac{x_{1}t}{z},\,\,\hat
u=\frac{x_{2}u}{z},\,\, z=-\frac{x_{1}t+x_{2}u}{x_{1}x_{2}s}.
\end{equation}
$D_{g}^{\pi}(z)=D_{g}^{\pi^{+}}(z)=D_{g}^{\pi^{-}}(z)$ represents
the gluon fragmentation function into a meson containing a gluon of
the same flavor. In the leading twist subprocess, $\pi$ meson is
indirectly emitted from the gluon with fractional momentum $z$. The
$\delta$ function may be expressed in terms of the parton kinematic
variables, and the $z$ integration may then be done. The final form
for the cross section is
$$
\Sigma_{M}^{LT}\equiv
E\frac{d\sigma}{d^3p}=\sum_{q}\int_{x_{1min}}^{1} dx_1
\int_{x_{2min}}^{1} dx_2 G_{{q_{1}}/{h_{1}}}(x_{1})
G_{{q_{2}}/{h_{2}}}(x_{2})D_{g}^{\pi}(z) \times
$$
\begin{equation}
\frac{1}{\pi z}\frac{d\sigma}{d\hat t}(q\bar{q} \to
g\gamma)=\sum_{q}\int_{x_{1min}}^{1} dx_1 \int_{x_{2min}}^{1} dx_2
\frac{x_{1}G_{{q_{1}}/{h_{1}}}(x_{1})
sx_{2}G_{{q_{2}}/{h_{2}}}(x_{2})}{-(x_{1}t+x_{2}u)}\frac{D_{g}^{\pi}(z)}{\pi}\frac{d\sigma}{d\hat
t}(q\bar{q} \to g\gamma).
\end{equation}

\section{NUMERICAL RESULTS AND DISCUSSION}\label{results}

In this section, the numerical results for higher twist effects on
the dependence of the chosen meson wave functions in the process $pp
\to \pi^{+}(or\,\, \pi^{-})\gamma$ are discussed. We have calculated
the dependence on the pion wave functions for the high twist
contribution to the  large-$p_T$ pion production cross section in
the proton-proton collision . In the calculations, the asymptotic
$\Phi_{asy}$, Chernyak-Zhitnitsky $\Phi_{CZ}$, two other pion model
functions, $\Phi_{P_2}$, $\Phi_{P_3}$ and also, the pion wave
function, from which two non-trivial Gegenbauer coefficients $a_2$
and $a_4$ have been extracted from the CLEO data on the
$\pi^{0}\gamma$ transition form factor have been used[20]. In the
ref.[20], authors have used the QCD light-cone sum rules approach
and included into their analysis the NLO perturbative and twist-four
corrections. For the high twist subprocess, we take $q_1+\bar{q}_{2}
\to (q_1\bar{q}_2)+\gamma$ and we have extracted the following four
high twist subprocess $u\bar{d} \to \pi^{+}\gamma,$ $\bar{d}u \to
\pi^{+}\gamma$, $\bar{u}d \to \pi^{-}\gamma$, $d\bar{u} \to
\pi^{-}\gamma$ contributing to $pp\to \pi^{+}(or\,\,\pi^{-})\gamma$
cross sections. For the dominant leading twist subprocess for the
pion production, we take the quark-antiquark annihilation $q\bar{q}
\to g\gamma$, in which the $\pi$ meson is indirectly emitted from
the gluon. As an example for the quark distribution function inside
the proton, the MRST2003c package [25] has been used. The gluon
fragmentation function has been taken from [26]. The other problems
dealth with are the choice of the QCD scale parameter $\Lambda$ and
the number of the active quark flavors $n_f$. The high twist
subprocesses probe the meson wave functions over a large range of
$Q^2$ squared momentum transfer, carried by the gluon. Therefore,
phenomenologically,  in the given diagram in Fig 1, for the
center-of-mass energy $\sqrt s=63 GeV$, we take $Q^2=p_{T}^2$.
However, for the center-of-mass  energy $\sqrt s=630 GeV$, we take
$Q_{1}^2=(x_1-1){\hat u}$, $Q_{2}^2=-x_1\hat t$ , which we have
obtained directly from the high twist subprocesses diagrams. The
same $Q^2$ has been used as an argument of $\alpha_s(Q^2)$ in the
calculation of each diagram. The results of our numerical
calculations are plotted in Figs.3-18. Figs.3-5 show the dependence
of the differential cross sections of the high twist
$\Sigma_{\pi^{+}}^{HT}$, leading twist plus high twist
$\Sigma_{\pi^{+}}^{tot}$=$\Sigma_{\pi^{+}}^{LT}$+$\Sigma_{\pi^{+}}^{HT}$
and ratio $R=\Sigma_{\pi^{+}}^{HT}$/$\Sigma_{\pi^{+}}^{LT}$ as a
function of the pion transverse momentum $p_T$ for five different
meson wave functions. As shown in Figs.3-4, the high twist
differential cross section is monotonically decreasing with an
increase in the transverse momentum of the pion. As seen from
Figs.3-4, in all wave functions of the mesons, the dependencies of
the high twist cross sections on the $p_T$ transverse momentum of
the pion demonstrate the same behavior. On the other hand, the
higher twist corrections are very sensitive to the choice of the
pion wave function. We should note that the magnitude of the high
twist cross section in the pion wave function $\Phi_{CLEO}(x,Q^2)$
case is very close to the asymptotic wave function $\Phi_{asy}(x)$
case. In Fig.5, the ratio
$R=\Sigma_{\pi^{+}}^{HT}$/$\Sigma_{\pi^{+}}^{LT}$ is plotted at
$y=0$ as a function of the pion transverse momentum $p_T$ for the
different pion wave functions. First of all, it is seen that the
values of $R$ for fixed $y$ and $\sqrt s$ depend on the choice of
the pion wave function. Also, the distinction between
$R(\Phi_{asy}(x))$ with $R(\Phi_{CLEO}(x,Q^2))$,
$R(\Phi_{CZ}(x,Q^2))$, $R(\Phi_{P_2}(x,Q^2))$ and
$R(\Phi_{P_3}(x,Q^2))$ have been calculated. We have found that the
distinction $R(\Phi_{asy}(x))$ and $R(\Phi_{CLEO}(x,Q^2))$ is small,
whereas a distinction  between $R(\Phi_{asy}(x))$ with
$R(\Phi_{CZ}(x,Q^2))$, $R(\Phi_{P_2}(x,Q^2))$ and
$R(\Phi_{P_3}(x,Q^2))$ is significant. For example, in the case of
$\sqrt s=63 GeV$, $y=0$, the distinction between $R(\Phi_{asy}(x))$
with $R(\Phi_{i}(x,Q^2))$ $(i=CLEO, CZ, P_2, P_3)$ is shown in Table
\ref{table1}.

Thus, the distinction between $R(\Phi_{asy}(x))$ and
$R(\Phi_{CLEO}(x,Q^2))$ is maximum at $p_T=6GeV/c$, but the
distinction between $R(\Phi_{asy}(x))$ with $R(\Phi_{CZ}(x,Q^2))$,
$R(\Phi_{P_2}(x,Q^2))$, $R(\Phi_{P_3}(x,Q^2))$ is maximum at $p_T=2
GeV/c$ and decreases with an increase in $p_T$. Such a behavior  of
$R$ may  be explained by reducing all moments of the pion model wave
functions to those of $\Phi_{asy}(x)$ for high $Q^2$. In Fig.6, we
show the difference of the
$\Delta_{\pi}^{HT}$=$\Sigma_{\pi^{+}}^{HT}$-$\Sigma_{\pi^{-}}^{HT}$
high twist cross section as a function of the pion transverse
momentum $p_T$ for five different pion wave functions. As seen from
Fig.6, when the transverse momentum of the pion is increasing, the
difference $\Delta_{\pi}^{HT}$ high twist cross section is
monotonically decreasing. As shown in Fig.6, the difference
$\Delta_{\pi}^{HT}$ high twist cross section for pion wave function
$\Phi_{CLEO}(x,Q^2)$ is very close to the asymptotic wave function
$\Phi_{asy}(x)$. In Fig.7, we show the difference
$\Delta_{\pi}^{LT}$=$\Sigma_{\pi^{+}}^{LT}$-$\Sigma_{\pi^{-}}^{LT}$
leading twist and sum of differences leading and high twist
$\Delta_{\pi}^{tot}$=$\Delta_{\pi}^{LT}$+$\Delta_{\pi}^{HT}$  cross
section for five pion wave functions as a function of the pion
transverse momentum $p_T$. As in Fig.6, the $\Delta_{\pi}^{LT}$ and
$\Delta_{\pi}^{tot}$ are monotonically decreasing when the
transverse momentum $p_T$ of the pion is increasing. But, the
distinction between the difference $\Delta_{\pi}^{LT}$ leading and
$\Delta_{\pi}^{tot}$=$\Delta_{\pi}^{LT}$+$\Delta_{\pi}^{HT}$ sum of
the difference leading and the high twist cross section is not
evident. In Fig.8, the ratio
$r={\Delta_{\pi}^{HT}}/{\Delta_{\pi}^{LT}}$ is plotted at $y=0$ as a
function of the pion transverse momentum $p_T$ for five pion wave
functions. As shown in Fig.8, the values of $r$ for fixed $y$ and
$\sqrt s$ depend on the choice of pion wave function as in Fig.5.
Also, we have calculated the distinction between $r(\Phi_{asy}(x))$
with $r(\Phi_{CLEO}(x,Q^2))$, $r(\Phi_{CZ}(x,Q^2))$,
$r(\Phi_{P_2}(x,Q^2))$ and $r(\Phi_{P_3}(x,Q^2))$. For example, in
the case of $\sqrt s=63GeV$, $y=0$  the distinction between
$r(\Phi_{asy}(x))$ with $r(\Phi_{i}(x,Q^2))$ (i=CLEO, CZ, $P_2$,
$P_3$) is presented in Table \ref{table2}.

We have obtained very interesting results. The calculations show
that the ratio $R(\Phi_{i}(x,Q^2))$/$R(\Phi_{asy}(x))$, (i=CLEO, CZ,
$P_2$, $P_3$) of all the transverse momentum $p_T$ of the pion is
equal to the ratio $r(\Phi_{i}(x,Q^2)$/$r(\Phi_{asy}(x))$, (i=CLEO,
CZ, $P_2$, $P_3$). In Figs.9-10, we have depicted
$\Sigma_{\pi^{+}}^{HT}$ and
$\Delta_{\pi}^{HT}$=$\Sigma_{\pi^{+}}^{HT}$-$\Sigma_{\pi^{-}}^{HT}$
as a function of the rapidity $y$ of the pion at $\sqrt s=63GeV$ and
$p_T=5 GeV/c$. As we are  now in the high energy region, the change
of the rapidity to determine these relations is given by $-ln(\sqrt
s/p_T)\leq y \leq ln(\sqrt s/p_T)$. At $\sqrt s=63 GeV$ and
$p_T=5GeV/c$, the pion rapidity lies in the region $-2.52\leq
y\leq2.52$. As seen from Figs.9-10, in the region ($-2.52\leq y\leq
0.85$), the high twist cross section for asymptotic $\Phi_{asy}(x)$
and $\Phi_{CLEO}(x,Q^2)$ wave functions increase  and it has a
maximum approximately at one point $y=0.85$. After that, the cross
sections monotonically decrease with an increase in the $y$ rapidity
of the pion. As seen from Figs.9-10, the high twist
$\Sigma_{\pi^{+}}^{HT}$ and difference high twist
$\Delta_{\pi}^{HT}$ cross sections are very sensitive to the choice
of meson wave functions. Also, as shown in Figs.9-10, in the regions
$-2.52\leq y\leq-1$ and $1\leq y\leq2.52$, the high twist cross
section for the $\Phi_{CLEO}(x,Q^2)$ is very close to the
$\Phi_{asy}(x)$ wave function case. We have also carried out
comparative calculations in the center-of-mass energy  $\sqrt s=630
GeV$. Figs.11-13 show the dependence of the differential cross
sections of the high twist $\Sigma_{\pi^{+}}^{HT}$, leading twist
plus high twist
$\Sigma_{\pi^{+}}^{tot}$=$\Sigma_{\pi^{+}}^{LT}$+$\Sigma_{\pi^{+}}^{HT}$
and ratio $R=\Sigma_{\pi^{+}}^{HT}$/$\Sigma_{\pi^{+}}^{LT}$ as a
function of the pion transverse momentum $p_T$ for five different
meson wave functions. As shown in Figs.11-12, the high twist
differential cross section is monotonically decreasing when the
transverse momentum of the pion is increasing. As seen from
Figs.11-12, for all wave functions of mesons, the dependencies of
the high twist cross sections on the transverse momentum of the pion
demonstrate the same behavior. But, the higher twist corrections are
very sensitive to the choice of the pion wave function. We should
note that the magnitude of the high twist cross section for the pion
wave function $\Phi_{CLEO}(x,Q^2)$ is very close to the asymptotic
wave function $\Phi_{asy}(x)$ case. In Fig.13, the ratio
$R=\Sigma_{\pi^{+}}^{HT}$/$\Sigma_{\pi^{+}}^{LT}$ is plotted at
$y=0$ as a function of the pion transverse momentum $p_T$ for the
different pion wave functions. First of all, it may be seen that the
values of $R$ for fixed $y$ and $\sqrt s$ depend on the choice of
the pion wave function. As in center-of-mass energy $\sqrt s=63GeV$,
we have also calculated the difference between $R(\Phi_{asy}(x))$
and four other wave functions \emph{i.e.} $R(\Phi_{CLEO}(x,Q^2)),$
$R(\Phi_{CZ}(x,Q^2)),$ $R(\Phi_{P_2}(x,Q^2)),$ and
$R(\Phi_{P_3}(x,Q^2))$. We have found that the distinction between
$R(\Phi_{asy}(x))$ and $R(\Phi_{CLEO}(x,Q^2))$ is very small,
whereas the distinction between $R(\Phi_{asy}(x))$ and
$R(\Phi_{CZ}(x,Q^2))$, $R(\Phi_{P_2}(x,Q^2))$ and
$R(\Phi_{P_3}(x,Q^2))$ is significant. In order to demonstrate this,
in the case of $\sqrt s=630 GeV$, $y=0$ the distinction between
$R(\Phi_{asy}(x))$ with $R(\Phi_{i}(x,Q^2))$ (i=CLEO, CZ, $P_2$,
$P_3$)  is shown in Table \ref{table3}.

Thus, the distinction between $R(\Phi_{asy}(x))$ and
$R(\Phi_{CLEO}(x,Q^2))$ is maximum at $p_T=60GeV/c$, but the
distinction between $R(\Phi_{asy}(x))$ with $R(\Phi_{CZ}(x,Q^2))$,
$R(\Phi_{P_2}(x,Q^2))$, $R(\Phi_{P_3}(x,Q^2))$ is maximum at $p_T=20
GeV/c$ and decreases with an increase in $p_T$. In Fig.14, we have
shown the difference
$\Delta_{\pi}^{HT}$=$\Sigma_{\pi^{+}}^{HT}$-$\Sigma_{\pi^{-}}^{HT}$
high twist cross section as a function of the pion transverse
momentum $p_T$, for five different pion wave functions. As seen from
Fig.14, as the transverse momentum of pion,  $p_T$, increases,  the
difference of the high twist cross section,  $\Delta_{\pi}^{HT}$,
monotonically decreases. As shown in Fig.14, $\Delta_{\pi}^{HT}$ for
the  $\Phi_{CLEO}(x,Q^2)$ pion wave function is very close to the
asymptotic wave function, $\Phi_{asy}(x)$. In Fig.15, we have shown
$\Delta_{\pi}^{LT}$=$\Sigma_{\pi^{+}}^{LT}$-$\Sigma_{\pi^{-}}^{LT}$
\emph{i.e.} the leading twist cross section, and
$\Delta_{\pi}^{tot}$=$\Delta_{\pi}^{LT}$+$\Delta_{\pi}^{HT}$
\emph{i.e.} the sum of the leading and the high twist cross
sections, for five pion wave functions, as a function of the pion
transverse momentum, $p_T$. As in Fig.7, the $\Delta_{\pi}^{LT}$ and
$\Delta_{\pi}^{tot}$ cross sections monotonically decrease when the
transverse momentum $p_T$ of the pion increases. But the distinction
between $\Delta_{\pi}^{LT}$ and
$\Delta_{\pi}^{tot}$=$\Delta_{\pi}^{LT}$+$\Delta_{\pi}^{HT}$ is not
evident. In Fig.16, the ratio
$r={\Delta_{\pi}^{HT}}/{\Delta_{\pi}^{LT}}$ is plotted at $y=0$ as a
function of the pion transverse momentum, $p_T$, for five pion wave
functions. As shown in Fig.16, the values of $r$ for fixed $y$ and
$\sqrt s$ depend on the choice of pion wave function similar to
Fig.8. We have also calculated the distinction between
$r(\Phi_{asy})$ and $r(\Phi_{CLEO}(x,Q^2)$, $r(\Phi_{CZ}(x,Q^2))$,
$r(\Phi_{P_2}(x,Q^2)$ and $r(\Phi_{P_3}(x,Q^2)$. As an example, in
the case of $\sqrt s=630 GeV$, $y=0$ the distinction between
$r(\Phi_{asy}(x))$ and $r(\Phi_{i}(x,Q^2))$ (i=CLEO, CZ, $P_2$,
$P_3$) is shown in Table  \ref{table4}.

As seen from calculations with increasing center-of-mass  energy
from $\sqrt s=63GeV$ to $\sqrt s=630GeV$, the distinction between
$R$ and $r$ decreases for all pion wave functions. In Figs.17-18, we
have depicted $\Sigma_{\pi^{+}}^{HT}$ and
$\Delta_{\pi}^{HT}$=$\Sigma_{\pi^{+}}^{HT}$-$\Sigma_{\pi^{-}}^{HT}$
as a function of the rapidity $y$ of pion at the $\sqrt s=630GeV$
and $p_T=50 GeV/c$. Since we are now in the high energy region, the
change of the rapidity to determine these relations is $-ln(\sqrt
s/p_T)\leq y \leq ln(\sqrt s/p_T)$. At $\sqrt s=630 GeV$ and
$p_T=50GeV/c$ the pion rapidity also lies in the region $-2.52\leq
y\leq2.52$. As seen from Figs.17-18, in the region ($-2.52\leq
y\leq0.85$), the high twist cross section for the asymptotic
$\Phi_{asy}(x)$ and $\Phi_{CLEO}(x,Q^2)$ wave functions increases
and it has a maximum approximately at one point $y=0.85$. After
that, the cross sections monotonically decrease with an increase in
the $y$ rapidity of the pion. As in Figs.9-10, in the regions
$-2.52\leq y\leq-1$ and $1\leq y\leq2.52$, the high twist cross
section for $\Phi_{CLEO}(x,Q^2)$ is very close to the
$\Phi_{asy}(x)$ wave function.

\section{Concluding Remarks}\label{conc}
In this work, we have calculated the higher twist contribution to
the large-$p_T$ pion production cross section to show  the
dependence on the chosen meson wave functions in the process $pp \to
\pi^{+}(or\,\,\pi^{-})\gamma$. In our calculations, we have used the
asymptotic $\Phi_{asy}$, Chernyak-Zhitnitsky $\Phi_{CZ}$, two other
pion model functions, $\Phi_{P_2}$, $\Phi_{P_3}$ and also, the pion
wave function, in which the coefficients $a_{2}$ and $a_{4}$ have
been extracted from the CLEO data on the $\pi^{0}\gamma$ transition
form factor used. For the high twist subprocess, we have taken
$q_1+\bar{q}_{2} \to(q_1\bar{q}_2)+\gamma$. We have extracted the
following four high twist subprocesses $u\bar{d} \to \pi^{+}\gamma$,
$\bar{d}u \to \pi^{+}\gamma$, $\bar{u}d \to \pi^{-}\gamma$,
$d\bar{u} \to \pi^{-}\gamma$, contributing to $pp \to
\pi^{+}(or\pi^{-})\gamma$ cross sections. As the dominant leading
twist subprocess for the pion production, we have taken the
quark-antiquark annihilation $q\bar{q} \to g\gamma$, where the $\pi$
meson is indirectly emitted from the gluon. The results of our
numerical calculations have been plotted in Figs.3-18. As shown in
Figs.3-4 and Figs.11-12, the high twist differential cross section
monotonically decrease when the transverse momentum of the pion
increases. As seen from Figs.3-4 and Figs.11-12 in all wave
functions of mesons, the dependencies of the high twist cross
sections on the $p_T$ transverse momentum of the pion demonstrate
the same behavior. But, the higher twist corrections are very
sensitive to the choice of the pion wave function. It should be
noted that the magnitude of the high twist cross section for the
pion wave function $\Phi_{CLEO}(x,Q^2)$ is very close to the
asymptotic wave function $\Phi_{asy}(x)$.

In Figs.5 and 13, the ratio
$R=\Sigma_{\pi^{+}}^{HT}$/$\Sigma_{\pi^{+}}^{LT}$ has been plotted
at $y=0$ as a function of the pion transverse momentum, $p_T$, for
the different pion wave functions. It may be observed that the
values of $R$ for fixed $y$ and $\sqrt s$ depend on the choice of
pion wave function. Within this context, we have also calculated the
distinction between $R(\Phi_{asy}(x))$ and $R(\Phi_{CLEO}(x,Q^2))$,
$R(\Phi_{CZ}(x,Q^2))$, $R(\Phi_{P_2}(x,Q^2))$ and
$R(\Phi_{P_3}(x,Q^2))$. We have ultimately found that the difference
between $R(\Phi_{asy}(x))$ and $R(\Phi_{CLEO}(x,Q^2))$ is small,
whereas a distinction  between $R(\Phi_{asy}(x))$ with
$R(\Phi_{CZ}(x,Q^2))$, $R(\Phi_{P_2}(x,Q^2))$ and
$R(\Phi_{P_3}(x,Q^2))$ is significant. In Figs.6 and 14, we have
shown the difference high twist cross section,
$\Delta_{\pi}^{HT}$=$\Sigma_{\pi^{+}}^{HT}$-$\Sigma_{\pi^{-}}^{HT}$,
as a function of the pion transverse momentum, $p_T$, for five
different pion wave functions. As seen from Figs.6 and 14 when the
transverse momentum of the pion increases, the  difference of the
high twist cross section, $\Delta_{\pi}^{HT}$, monotonically
decreases. As shown in Figs.6 and 14, $\Delta_{\pi}^{HT}$ in the
$\Phi_{CLEO}(x,Q^2)$ pion wave function is very close to that of the
asymptotic wave function $\Phi_{asy}(x)$.

In Figs.8 and 16, the ratio
$r={\Delta_{\pi}^{HT}}/{\Delta_{\pi}^{LT}}$ is plotted at $y=0$ as a
function of the pion transverse momentum, $p_T$, for five pion wave
functions. As shown in Figs.8 and 16, the values of $r$ for fixed
$y$ and $\sqrt s$ depend on the choice of the pion wave function
similar to Fig.5 and 13. We have also calculated the distinction
between $r(\Phi_{asy}(x))$ and $r(\Phi_{CLEO}(x,Q^2))$,
$r(\Phi_{CZ}(x,Q^2))$, $r(\Phi_{P_2}(x,Q^2))$ and
$r(\Phi_{P_3}(x,Q^2))$. For all transverse momentum $p_T$ of the
pion in the center-of-mass  energy $\sqrt s= 63GeV$ and also in
$630GeV$, we have obtained the following interesting relation:
$$
\frac{R(\Phi_{i}(x,Q^2))}{R(\Phi_{asy}(x))}=
\frac{r(\Phi_{i}(x,Q^2))}{r(\Phi_{asy}(x))},\,\, \,\,\,(i=CLEO, CZ,
P_2. P_3)
$$
In Figs.9-10 and Figs.17-18, we have depicted
$\Sigma_{\pi^{+}}^{HT}$ high twist and difference high twist cross
sections
$\Delta_{\pi}^{HT}$=$\Sigma_{\pi^{+}}^{HT}$-$\Sigma_{\pi^{-}}^{HT}$
as a function of the rapidity $y$ of the pion at $\sqrt s=63GeV,
630GeV$ and $p_T=5GeV/c, 50 GeV/c$, respectively. As we are now in
the high energy region, the change of the rapidity of this relation
may be expressed as folows: $-ln(\sqrt s/p_T)\leq y \leq ln(\sqrt
s/p_T)$. At $\sqrt s=63 GeV$, $p_T=5 GeV/c$ and $\sqrt s=630 GeV$,
$p_T=50 GeV/c$ the pion rapidity lies in the region $-2.52\leq
y\leq2.52$. As seen from Figs.9-10 and Figs.17-18 in the region
($-2.52\leq y\leq 0.85$), the high twist cross section for the
asymptotic $\Phi_{asy}(x)$ and $\Phi_{CLEO}(x,Q^2)$ wave functions
increases and it has a maximum approximately at one point $y=0.85$.
After that, the cross sections monotonically decrease with an
increase in the $y$ rapidity of the pion. As seen from Figs.9-10 and
Figs.17-18, the $\Sigma_{\pi^{+}}^{HT}$ and the $\Delta_{\pi}^{HT}$
cross sections are very sensitive to the choice of the meson wave
functions. Also, as shown in Figs.9, 10, 17 and 18 in the regions
$-2.52\leq y\leq-1$ and $1\leq y\leq2.52$ the high twist cross
section for the $\Phi_{CLEO}(x,Q^2)$ is very close to  the
asymptotic wave function $\Phi_{asy}(x)$ case. Our investigation
enables us to conclude that the high twist pion production cross
section in the proton-proton collisions depends on the form of the
pion model wave functions and may be used for their study. Further
investigations are needed in order to clarify the role of high twist
effects  in QCD.

\section*{Acknowledgments}
Two of authors, A. I.~Ahmadov and I.~Boztosun are grateful to
T\"{U}B\.{I}TAK Grant-2221 (BAYG) as well as T\"{U}B\.{I}TAK Grant:
TBAG-2398. One of us, A. I.~Ahmadov is also  grateful to NATO
Reintegration Grant-980779

\newpage
\begin{table}[h]
\begin{center}
\begin{tabular}{|c|c|c|c|c|c} \hline
$p_{T},GeV/c$ & $\frac{R(\Phi_{CLEO}(x,Q^2))}{R(\Phi_{asy}(x))}$ &
$\frac{R(\Phi_{CZ}(x,Q^2))}{R(\Phi_{asy}(x))}$ &
$\frac{R(\Phi_{P_2}(x,Q^2))}{R(\Phi_{asy}(x))}$&$\frac{R(\Phi_{P_3}(x,Q^2))}{R(\Phi_{asy}(x))}$\\
\hline
  2 & 1.496  & 8.435 & 10.569  &16.525 \\ \hline
  6 & 3.358  & 3.888 & 4.617 &3.001  \\ \hline
  20& 0.9476  & 0.382987 & 0.302073 & 0.133824 \\ \hline
\end{tabular}
\end{center}
\caption{The distinction between $R(\Phi_{asy}(x))$ with
$R(\Phi_{i}(x,Q^{2}))$  (i=CLEO, CZ, $P_{2}$, $P_{3}$) at c.m.
 energy $\sqrt s=63GeV$.} \label{table1}
\end{table}

\begin{table}[h]
\begin{center}
\begin{tabular}{|c|c|c|c|c|c}\hline
$p_{T},GeV/c$ & $\frac{r(\Phi_{CLEO}(x,Q^2))}{r(\Phi_{asy}(x))}$ &
$\frac{r(\Phi_{CZ}(x,Q^2))}{r(\Phi_{asy}(x))}$ &
$\frac{r(\Phi_{P_2}(x,Q^2))}{r(\Phi_{asy}(x))}$&$\frac{r(\Phi_{P_3}(x,Q^2))}{r(\Phi_{asy}(x))}$\\
\hline 2 & 1.496  & 8.435 & 10.568  &16.524 \\ \hline 6 & 3.358
&3.887 & 4.616 &3.001\\ \hline 20& 0.95  & 0.38315 &
0.302109&0.13384 \\ \hline
\end{tabular}
\end{center}
\caption{The distinction between $r(\Phi_{asy}(x))$ with
$r(\Phi_{i}(x,Q^{2}))$ (i=CLEO, CZ, $P_{2}$, $P_{3}$) at c.m. energy
$\sqrt s=63GeV$.} \label{table2}
\end{table}

\begin{table}[h]
\begin{center}
\begin{tabular}{|c|c|c|c|c|c}                  \hline
$p_{T},GeV/c$ & $\frac{R(\Phi_{CLEO}(x,Q^2))}{R(\Phi_{asy}(x))}$ &
$\frac{R(\Phi_{CZ}(x,Q^2))}{R(\Phi_{asy}(x))}$ &
$\frac{R(\Phi_{P_2}(x,Q^2))}{R(\Phi_{asy}(x))}$&$\frac{R(\Phi_{P_3}(x,Q^2))}{R(\Phi_{asy}(x))}$\\
\hline
  20 & 1.531  & 5.582 & 6.815  &9.63 \\ \hline
  60 & 2.202  & 2.591 & 2.961 &2.262  \\ \hline
  200& 0.926  & 0.605 & 0.544 & 0.43 \\ \hline
\end{tabular}
\end{center}
\caption{The distinction between $R(\Phi_{asy}(x))$ with
$R(\Phi_{i}(x,Q^{2}))$  (i=CLEO, CZ, $P_{2}$, $P_{3}$) at c.m.
energy $\sqrt s=630 GeV$.}\label{table3}
\end{table}

\begin{table}[h]
\begin{center}
\begin{tabular}{|c|c|c|c|c|c}\hline
$p_{T},GeV/c$ & $\frac{r(\Phi_{CLEO}(x,Q^2))}{r(\Phi_{asy}(x))}$ &
$\frac{r(\Phi_{CZ}(x,Q^2))}{r(\Phi_{asy}(x))}$ &
$\frac{r(\Phi_{P_2}(x,Q^2))}{r(\Phi_{asy}(x))}$&$\frac{r(\Phi_{P_3}(x,Q^2))}{r(\Phi_{asy}(x))}$\\
\hline
  20 & 1.531  & 5.583 & 6.817  &9.632 \\ \hline
  60 & 2.202  & 2.591 & 2.961 &2.262  \\ \hline
  200& 0.926  & 0.605 & 0.544 & 0.43 \\ \hline
\end{tabular}
\end{center}
\caption{The distinction between $r(\Phi_{asy}(x))$ with
$r(\Phi_{i}(x,Q^{2}))$ (i=CLEO, CZ, $P_{2}$, $P_{3}$)) at c.m.
energy $\sqrt s=630 GeV$.} \label{table4}
\end{table}

%\begin{center}
%{\bf FIGURE CAPTIONS \\}
%\end{center}
%\noindent
\begin{figure}[h]
\epsfxsize 20cm \centerline{\epsfbox{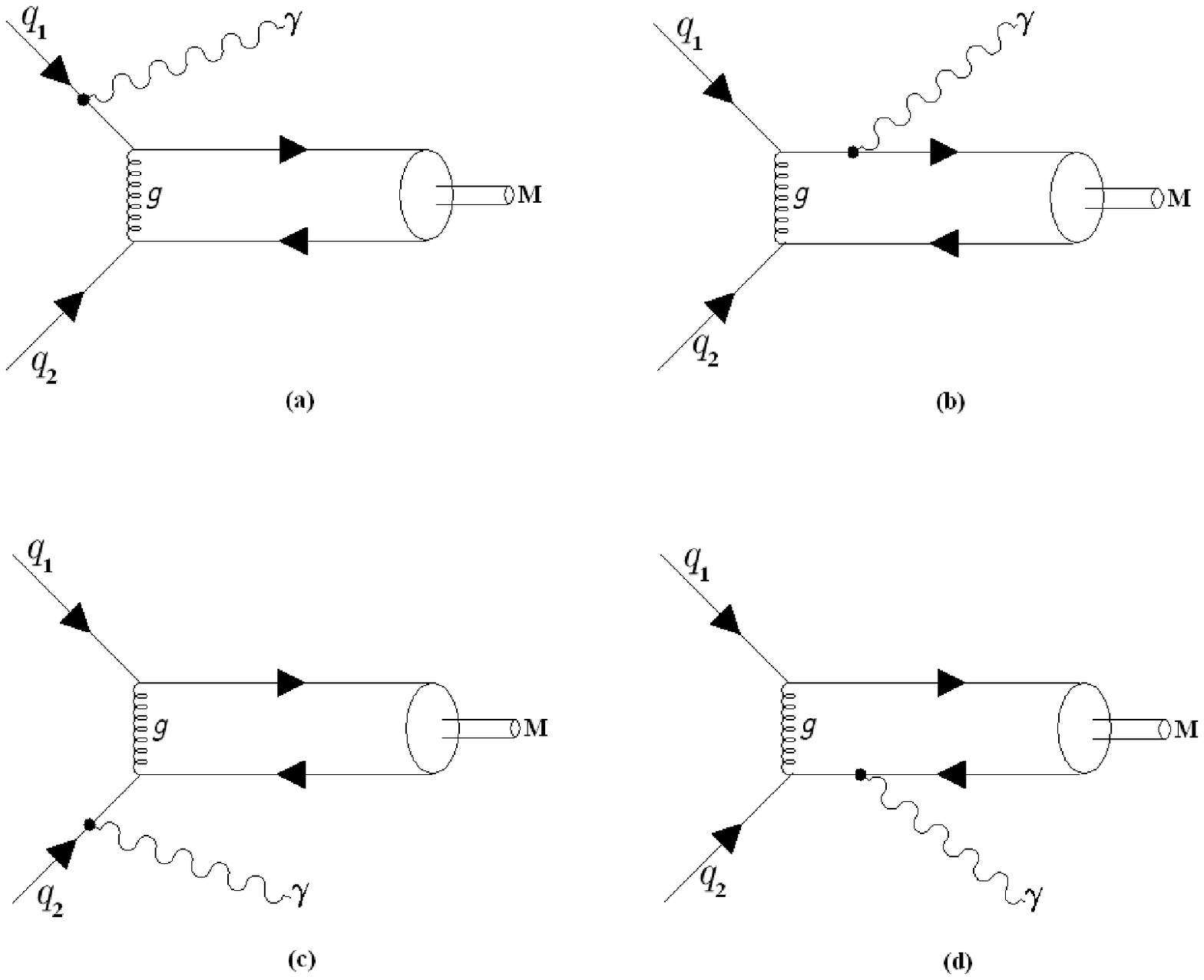}} \vskip-13cm
\caption{Feynman diagrams for the  high twist subprocess,
$q_1q_2\to\pi^{+}(or\pi^{-})\gamma$} \label{Fig1}
%\end{figure}
 \vskip-8.0cm
%\begin{figure}[h]
\epsfxsize 15.5cm \centerline{\epsfbox{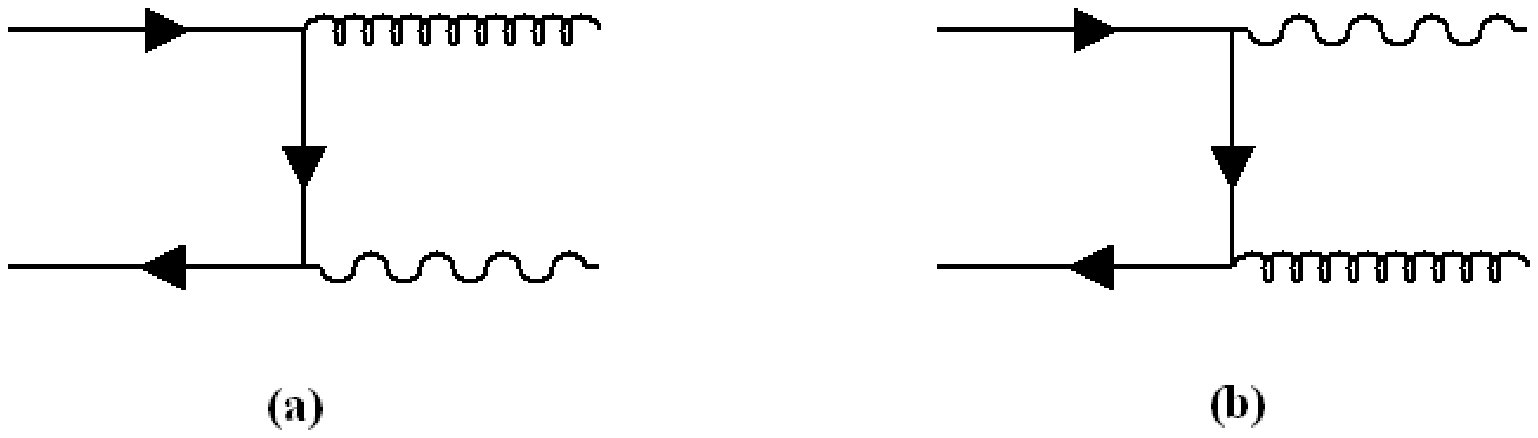}} \vskip-9.0cm
\caption{Feynman diagrams for the leading twist subprocess,
$q\overline{q}\to g\gamma, g\to M$.} \label{Fig2}
\end{figure}

\newpage

\begin{figure}[htb]
 \vskip-2.0cm\epsfxsize 21.5cm \centerline{\epsfbox{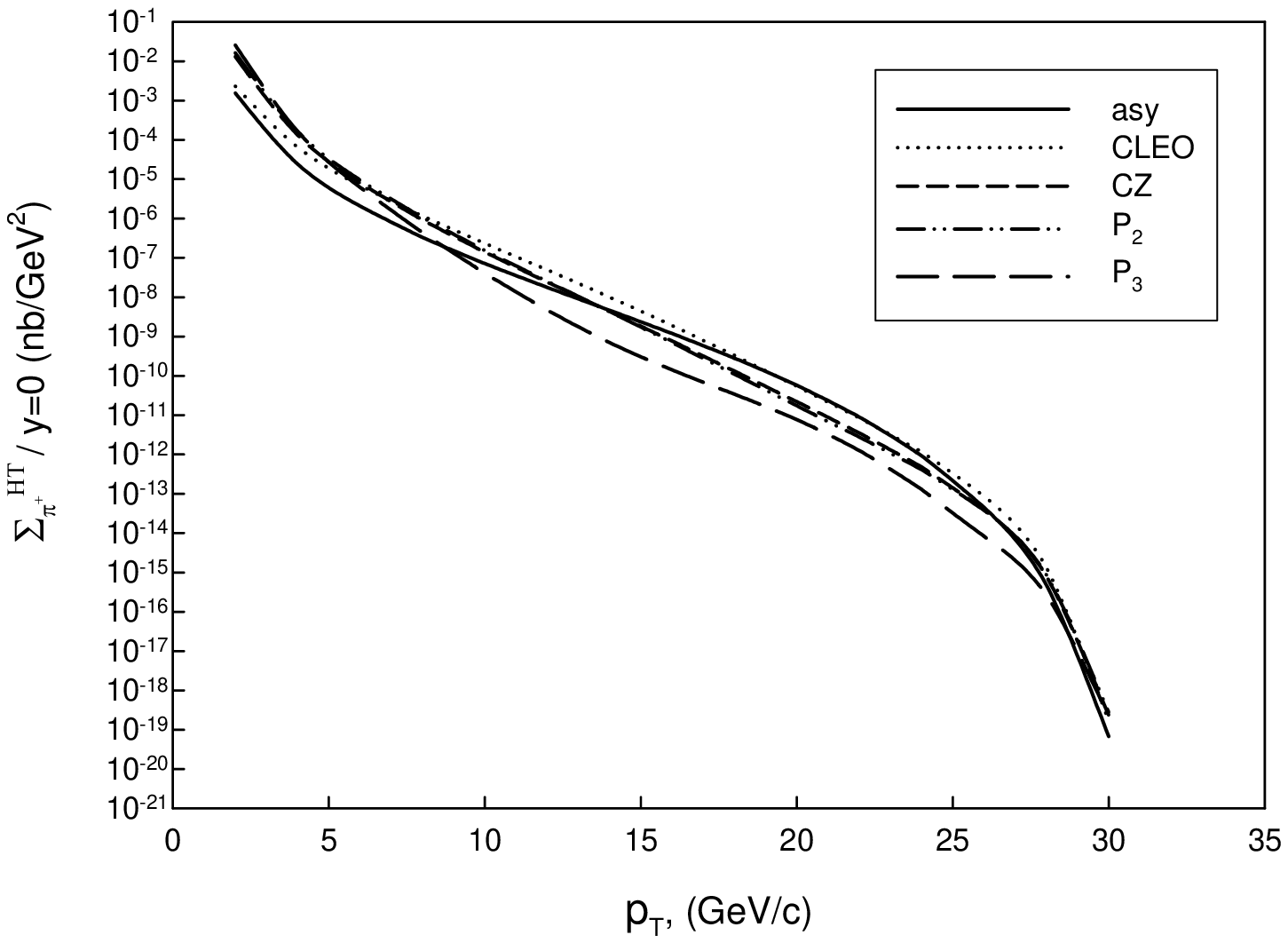}} \vskip-19.5cm
\caption{High twist $\pi^{+}$ production cross sections as a
function of the $p_T$ transverse momentum of the pion at the
c.m.energy $\sqrt s=63GeV$.} \label{Fig3}
%\end{figure}
 \vskip-1.0cm
%\begin{figure}[h]
\epsfxsize 21.5cm \centerline{\epsfbox{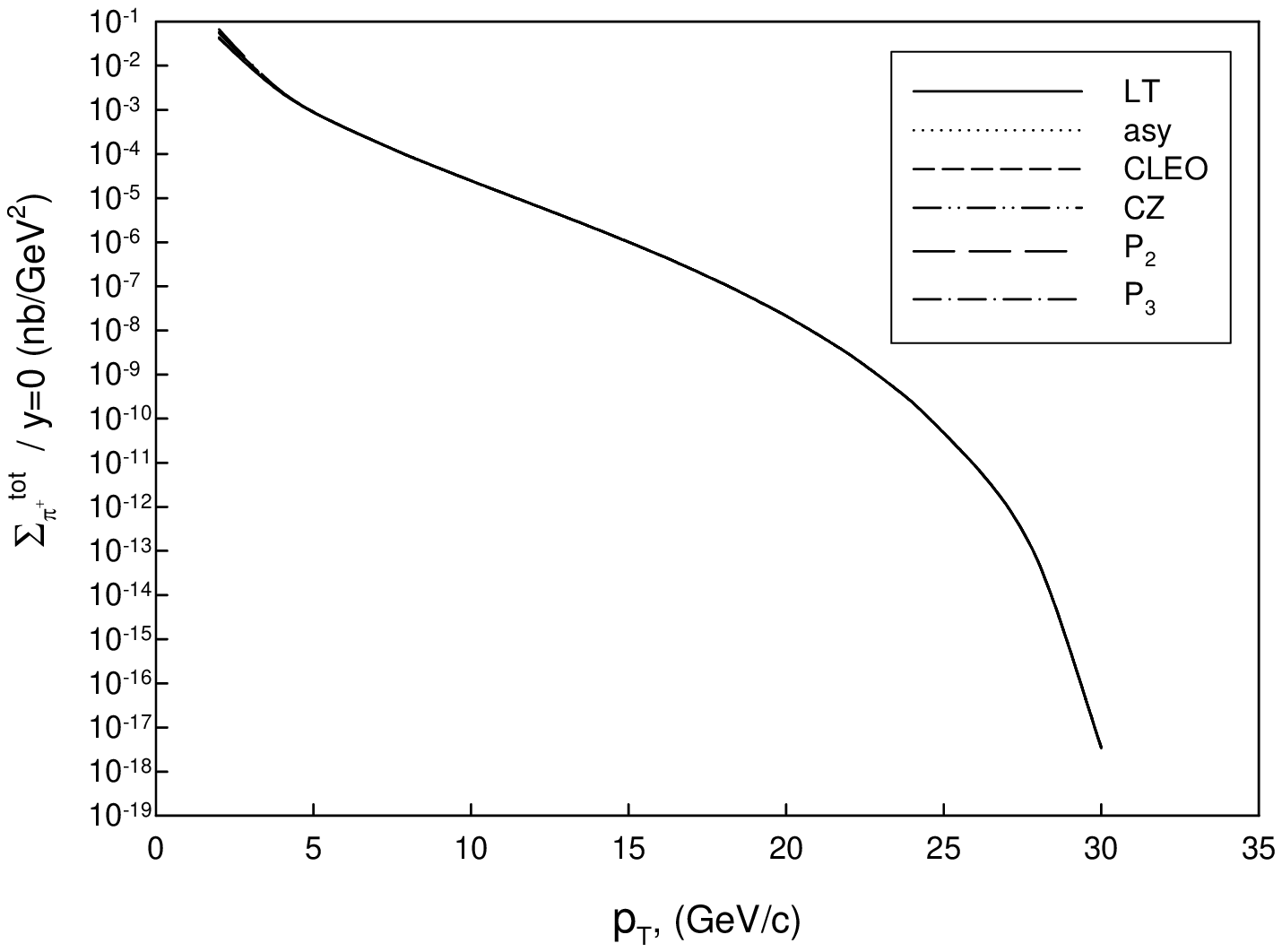}} \vskip-18.6cm
\caption{ The sum of the leading and the high twist $\pi^{+}$
production cross sections
$\Sigma_{\pi^{+}}^{tot}=\Sigma_{\pi^{+}}^{LT}+\Sigma_{\pi^{+}}^{HT}$
as a function of the $p_T$ transverse  momentum of the pion , at the
c.m. energy $\sqrt s=63GeV$.} \label{Fig4}
\end{figure}

\newpage

\begin{figure}[htb]
 \vskip-2.20cm\epsfxsize 21.5cm \centerline{\epsfbox{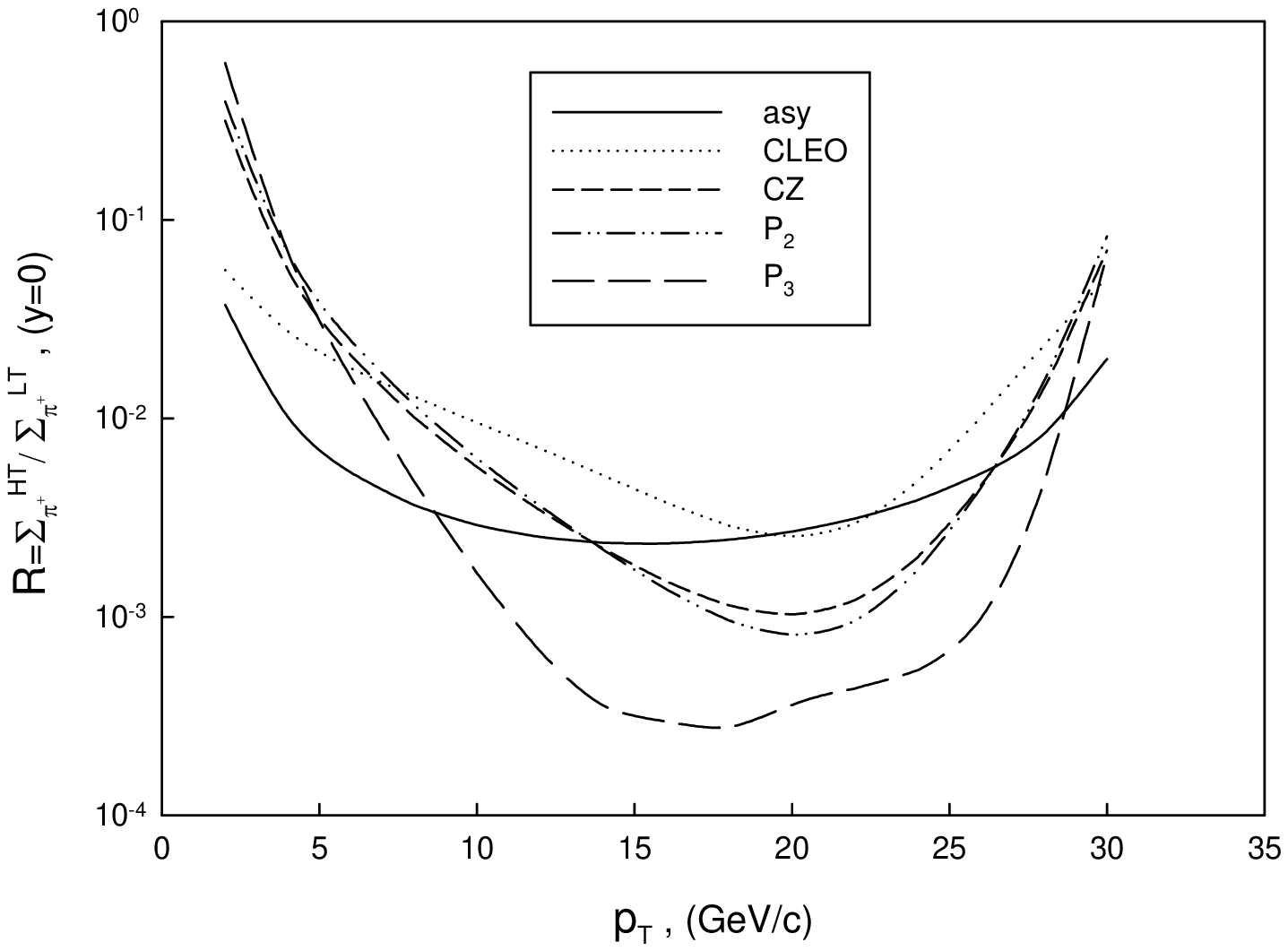}} \vskip-19.95cm
\caption{Ratio $R=\Sigma_{\pi^{+}}^{HT}$/$\Sigma_{\pi^{+}}^{LT}$,
where the leading and the high twist contributions are calculated
for the pion rapidity $y=0$ at the c.m. energy $\sqrt s=63GeV$, as a
function of the pion transverse momentum, $p_T$.} \label{Fig5}
%\end{figure}
 \vskip-0.0cm
%\begin{figure}[h]
\epsfxsize 21.5cm \centerline{\epsfbox{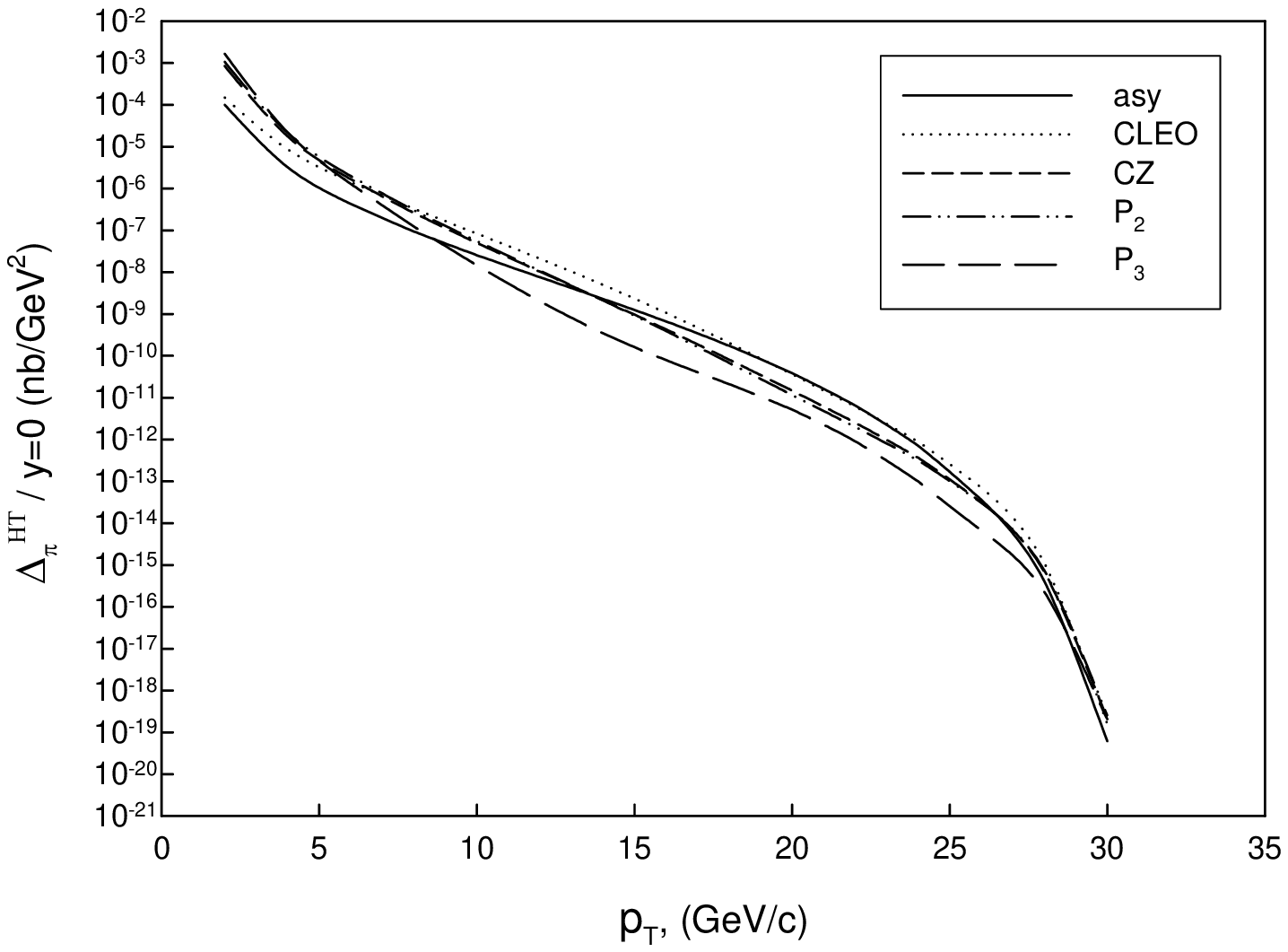}} \vskip-20.0cm
\caption{The difference of the high twist cross section,
$\Delta_{\pi}^{HT}$=$\Sigma_{\pi^{+}}^{HT}$-$\Sigma_{\pi^{-}}^{HT}$,
as a function of the  pion transverse momentum, $p_T$, at the c.m.
energy $\sqrt s=63GeV$.} \label{Fig6}
\end{figure}

\newpage

\begin{figure}[htb]
 \vskip-2.0cm\epsfxsize 21.2cm \centerline{\epsfbox{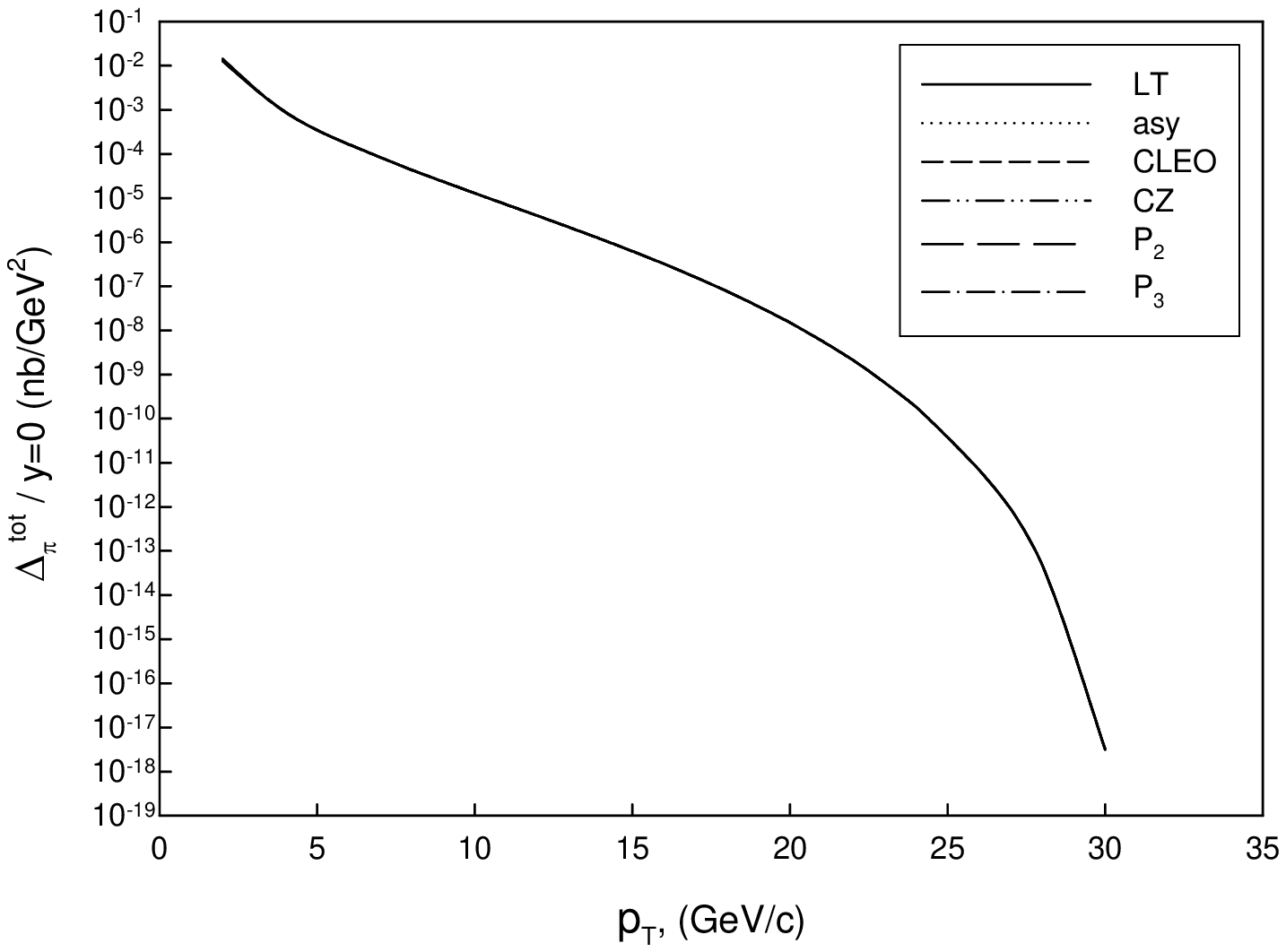}} \vskip-19.5cm
\caption{The sum of the difference leading and high twist  cross
sections
$\Delta_{\pi}^{tot}$=$\Delta_{\pi}^{LT}$+$\Delta_{\pi}^{HT}$ as a
function of the  pion transverse momentum $p_T$ at the c.m. energy
$\sqrt s=63GeV$.} \label{Fig7}
%\end{figure}
 \vskip-0.0cm\epsfxsize 21.2cm \centerline{\epsfbox{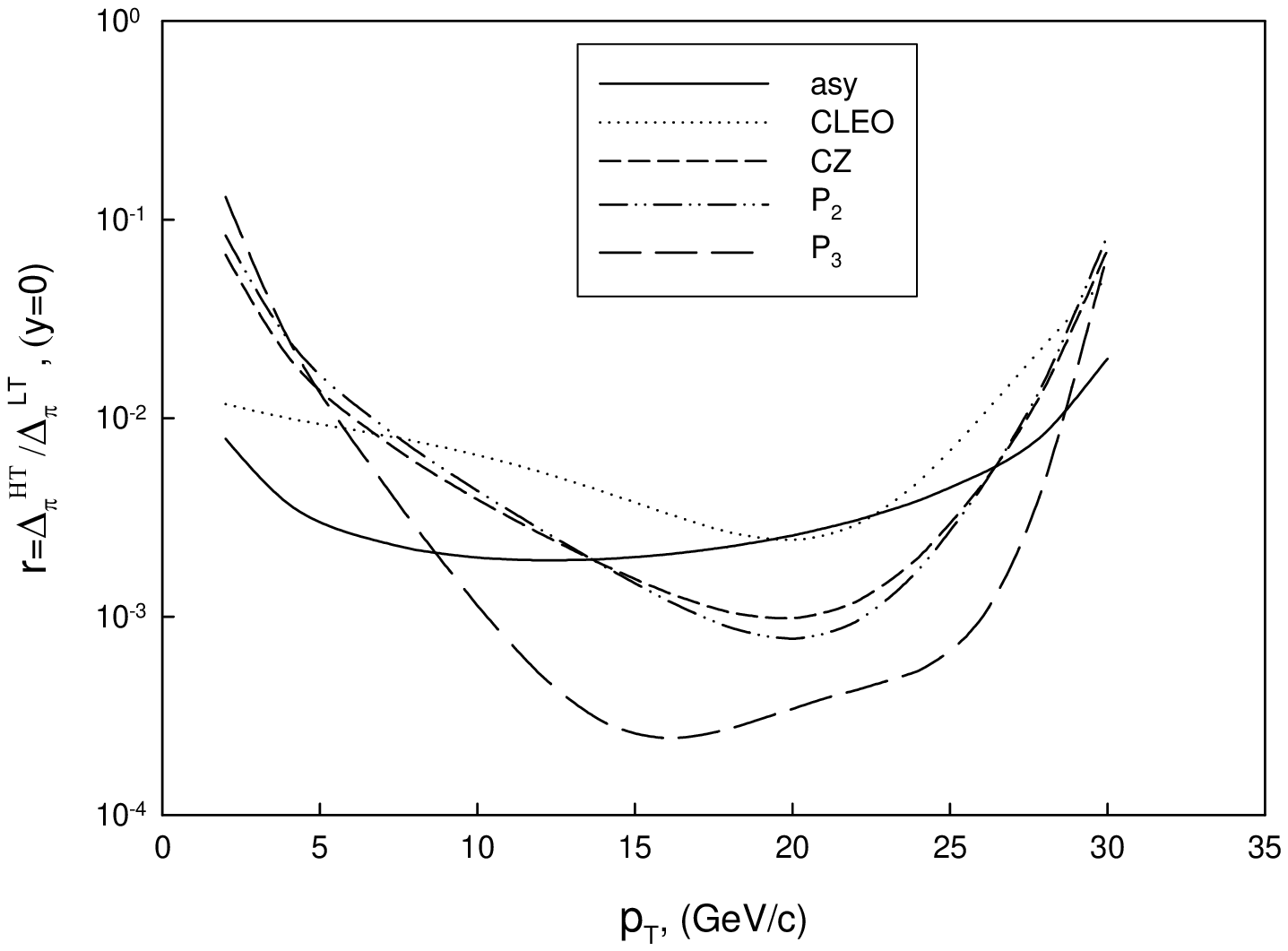}} \vskip-19.5cm
\caption{The ratio $r={\Delta_{\pi}^{HT}}/{\Delta_{\pi}^{LT}}$,
where the leading and the high twist contributions are calculated
for the pion rapidity $y=0$, at the c.m. energy $\sqrt s=63GeV$, as
a function of the pion transverse momentum $p_T$.} \label{Fig8}
\end{figure}

\newpage

\begin{figure}[htb]
 \vskip-2.0cm\epsfxsize 21.5cm \centerline{\epsfbox{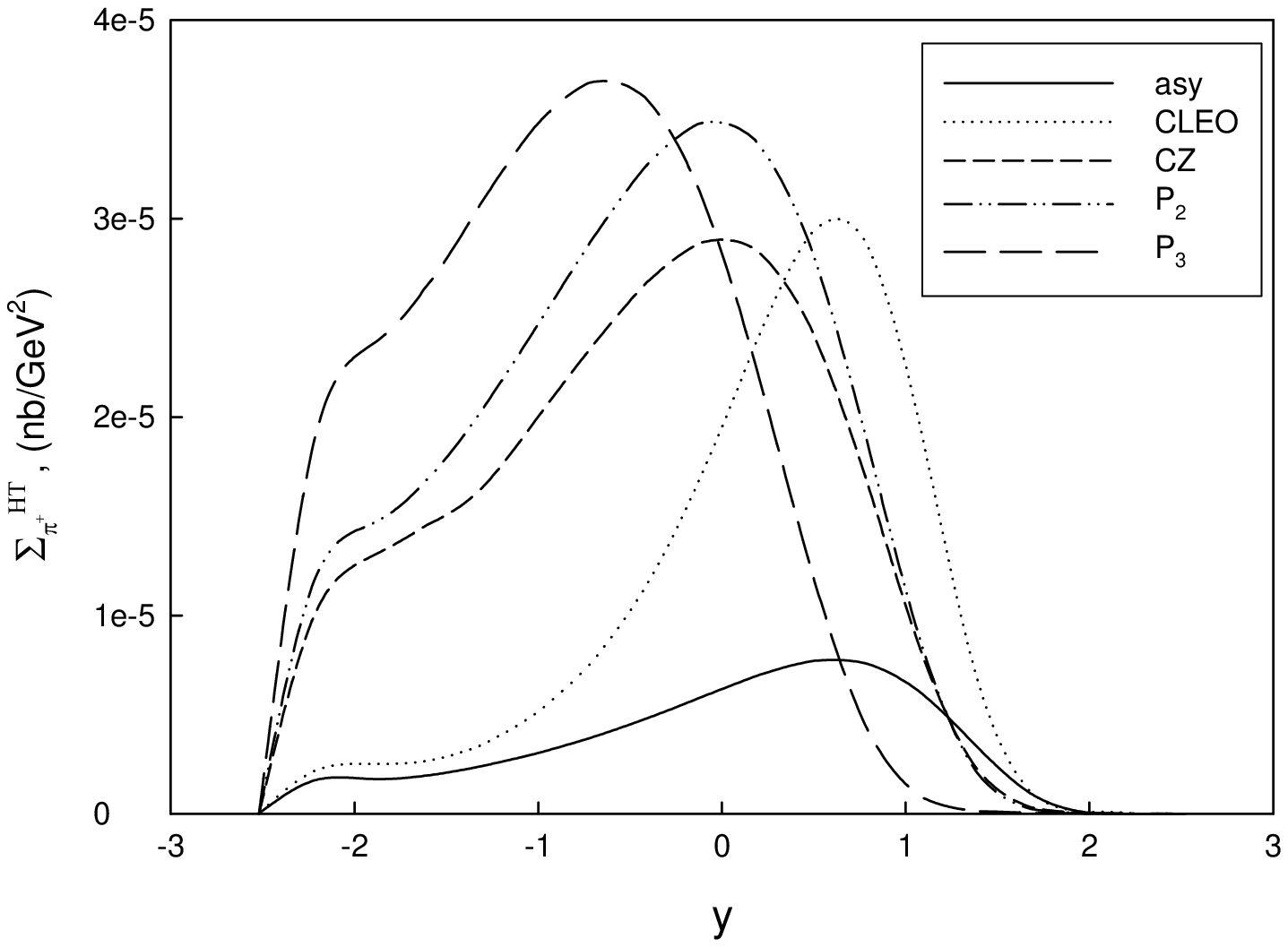}} \vskip-19.75cm
\caption{High twist $\pi^{+}$ production cross sections as a
function of the $y$ rapidity of the pion at the transverse momentum
of the pion $p_T=5 GeV$, at the c.m.energy $\sqrt s=63GeV$.}
\label{Fig9}
%\end{figure}
%\begin{figure}[h]
\epsfxsize 21.5cm \centerline{\epsfbox{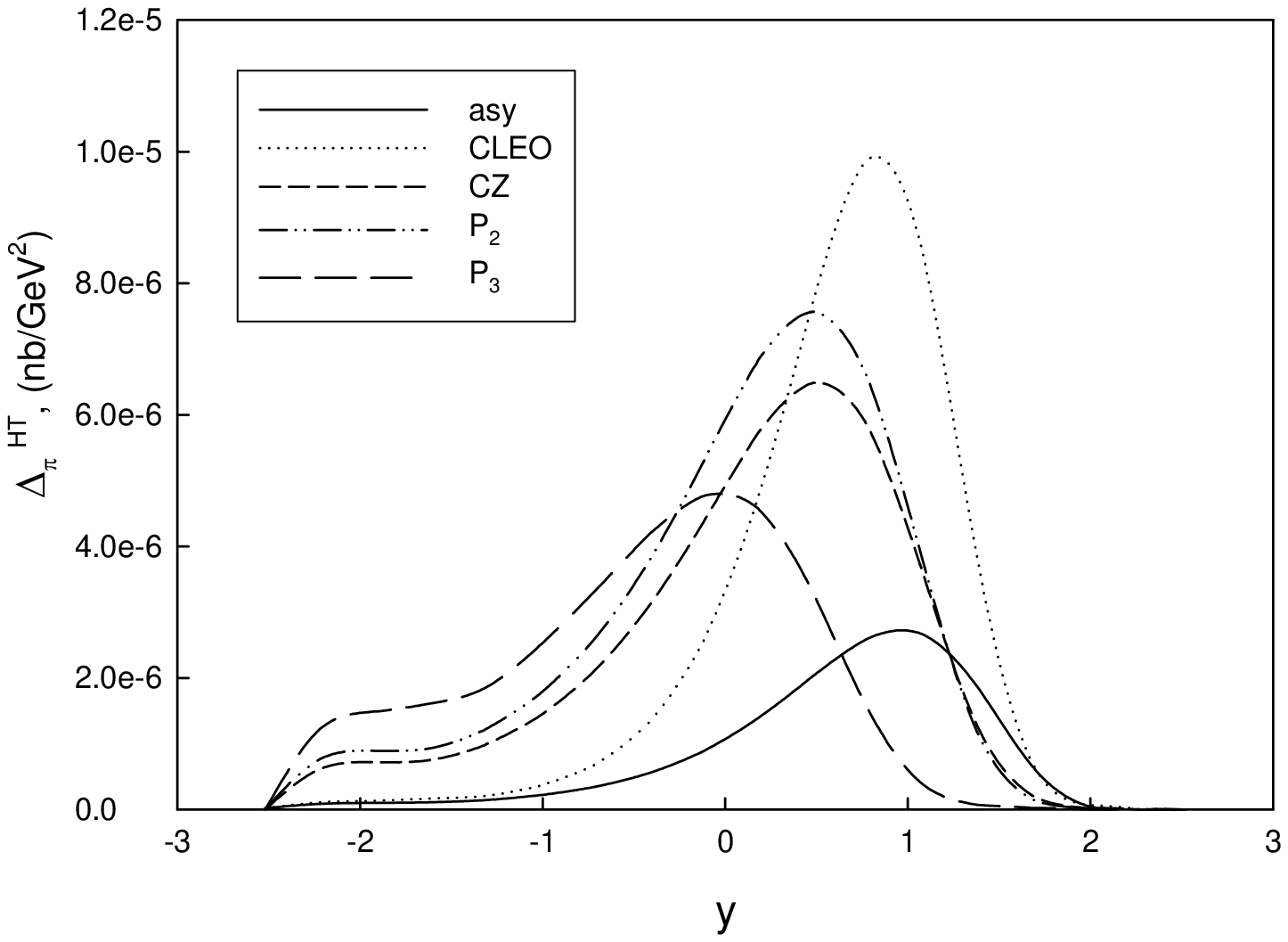}} \vskip-19.75cm
\caption{The difference of the high twist cross section,
$\Delta_{\pi}^{HT}$=$\Sigma_{\pi^{+}}^{HT}$-$\Sigma_{\pi^{-}}^{HT}$,
as a function of the $y$ rapidity of the pion at the  transverse
momentum of the pion $p_T=5 GeV$, at the c.m. energy $\sqrt
s=63GeV$.} \label{Fig10}
\end{figure}

\newpage

\begin{figure}[htb]
 \vskip-2.0cm\epsfxsize 21.5cm \centerline{\epsfbox{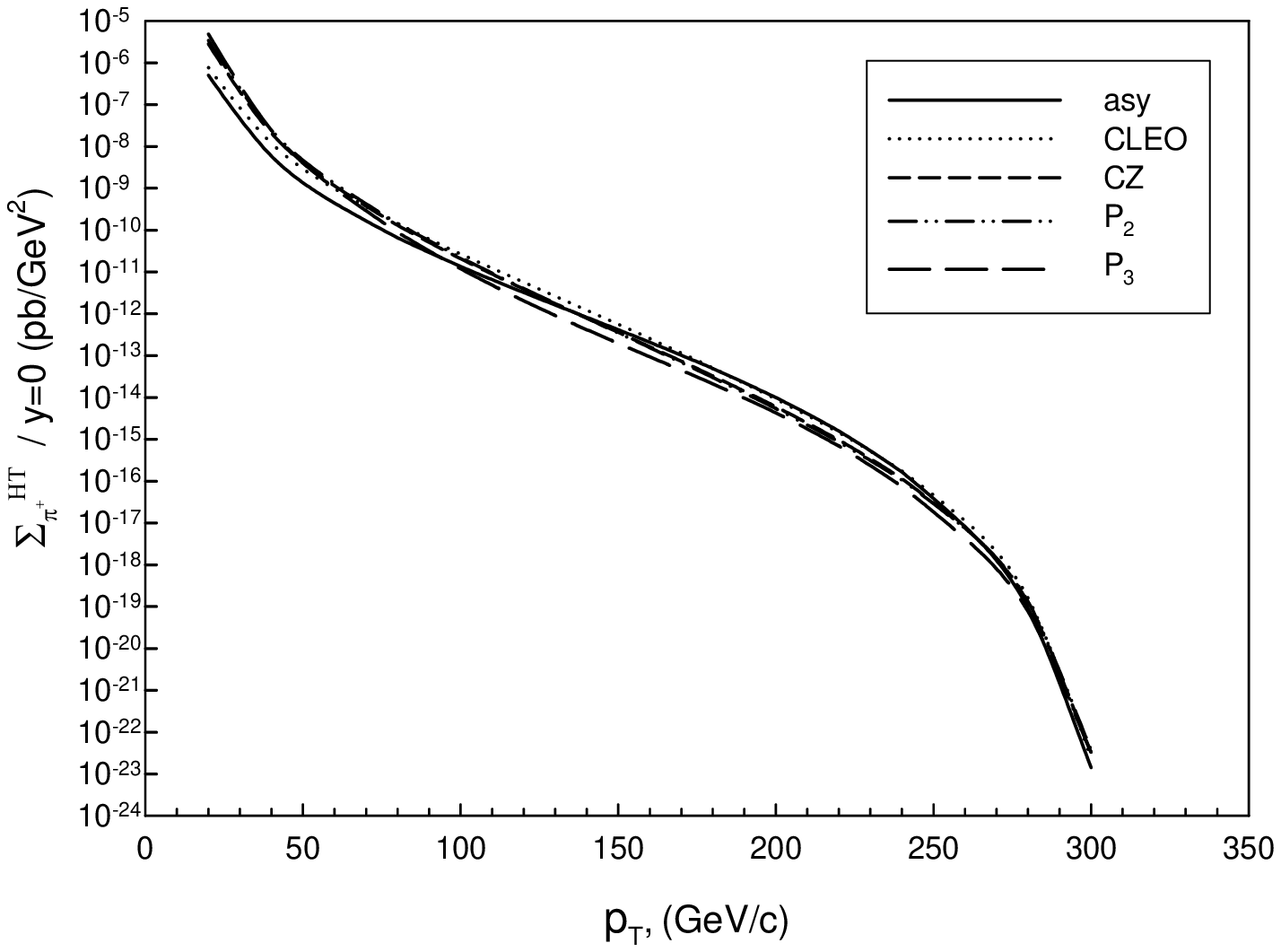}} \vskip-19.75cm
\caption{High twist $\pi^{+}$ production cross sections as a
function of the $p_T$ transverse momentum of the pion at the c.m.
energy $\sqrt s=630GeV$.} \label{Fig11}
%\end{figure}
%\begin{figure}[h]
\epsfxsize 21.5cm \centerline{\epsfbox{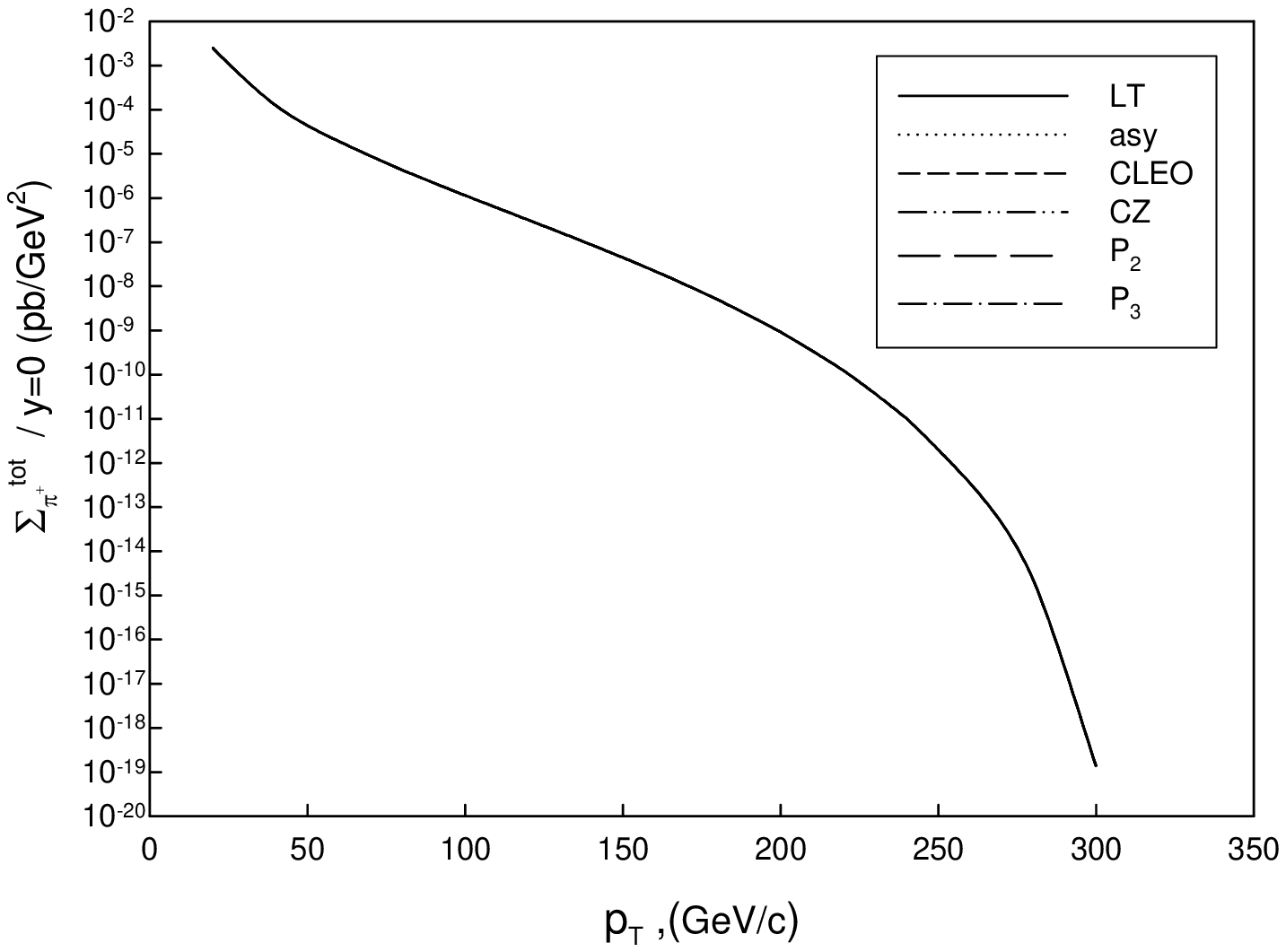}} \vskip-19.75cm
\caption{The sum of the leading and the high twist $\pi^{+}$
production cross sections
$\Sigma_{\pi^{+}}^{tot}=\Sigma_{\pi^{+}}^{LT}+\Sigma_{\pi^{+}}^{HT}$
as a  function of the $p_T$ transverse  momentum of the pion, at the
c.m. energy $\sqrt s= 630GeV$.} \label{Fig12}
\end{figure}

\newpage

\begin{figure}[htb]
 \vskip-2.2cm\epsfxsize 21.5cm \centerline{\epsfbox{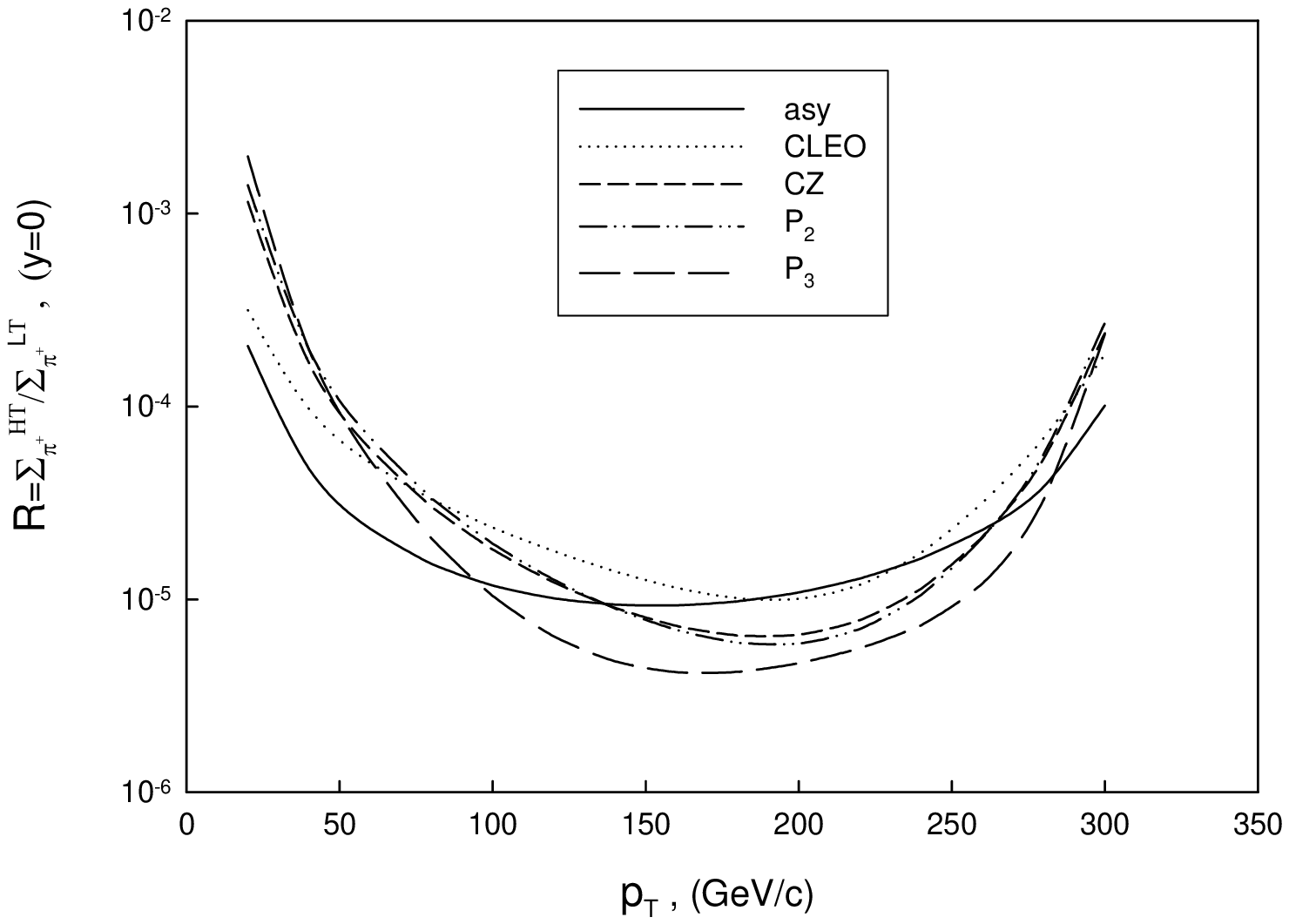}} \vskip-19.95cm
\caption{The ratio
$R=\Sigma_{\pi^{+}}^{HT}$/$\Sigma_{\pi^{+}}^{LT}$, where the leading
and the high twist contributions are calculated for the pion
rapidity $y=0$ at the c.m. energy $\sqrt s=630GeV$ as a function of
the pion transverse momentum $p_T$.} \label{Fig13}
%\end{figure}
%\begin{figure}[h]
\epsfxsize 21.5cm \centerline{\epsfbox{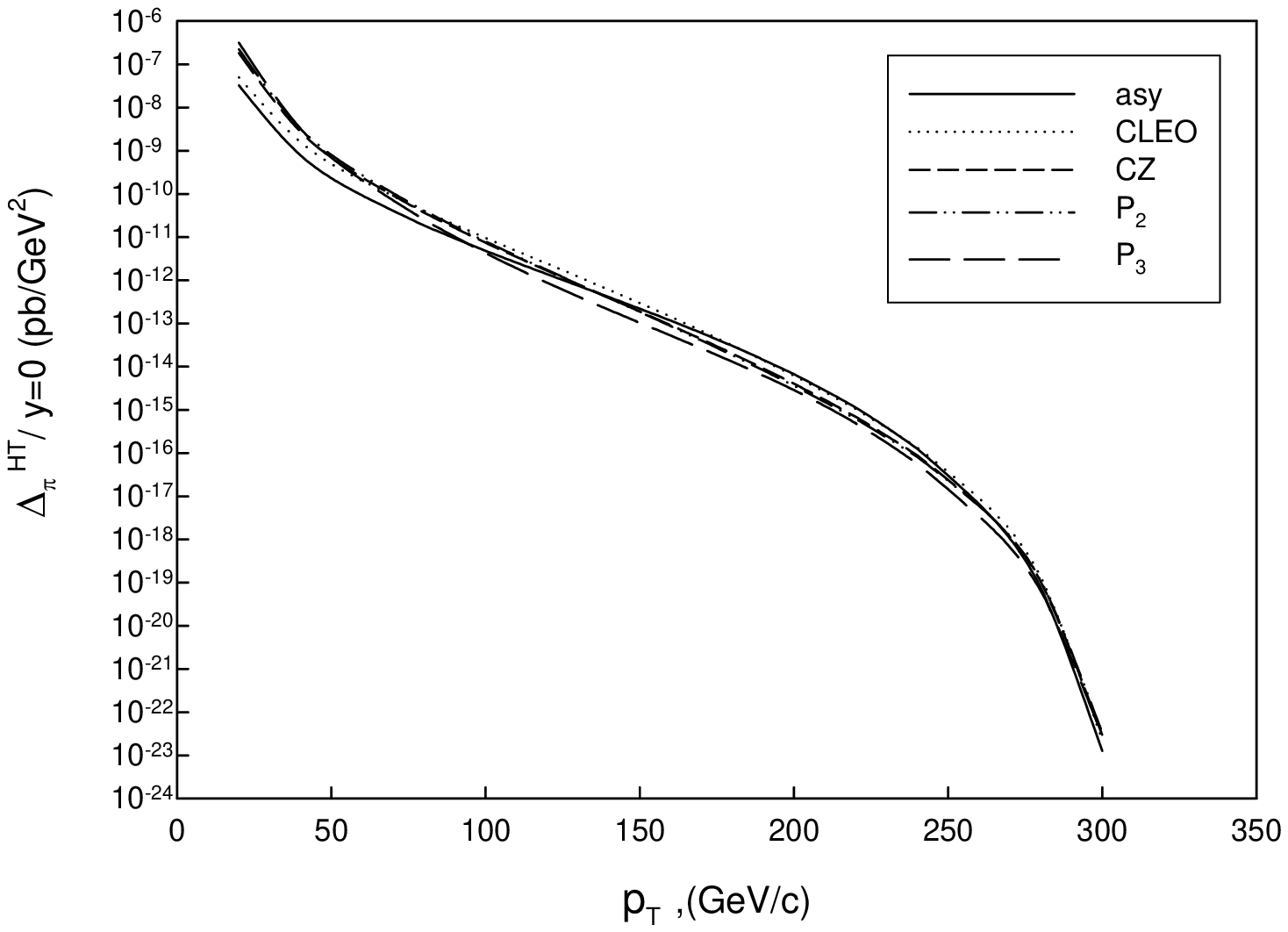}} \vskip-19.95cm
\caption{The difference of the high twist cross section,
$\Delta_{\pi}^{HT}$=$\Sigma_{\pi^{+}}^{HT}$-$\Sigma_{\pi^{-}}^{HT}$,
as a function of the  pion transverse momentum, $p_T$, at the c.m.
energy $\sqrt s=630GeV$.} \label{Fig14}
\end{figure}

\newpage

\begin{figure}[htb]
 \vskip-2.3cm\epsfxsize 21.5cm \centerline{\epsfbox{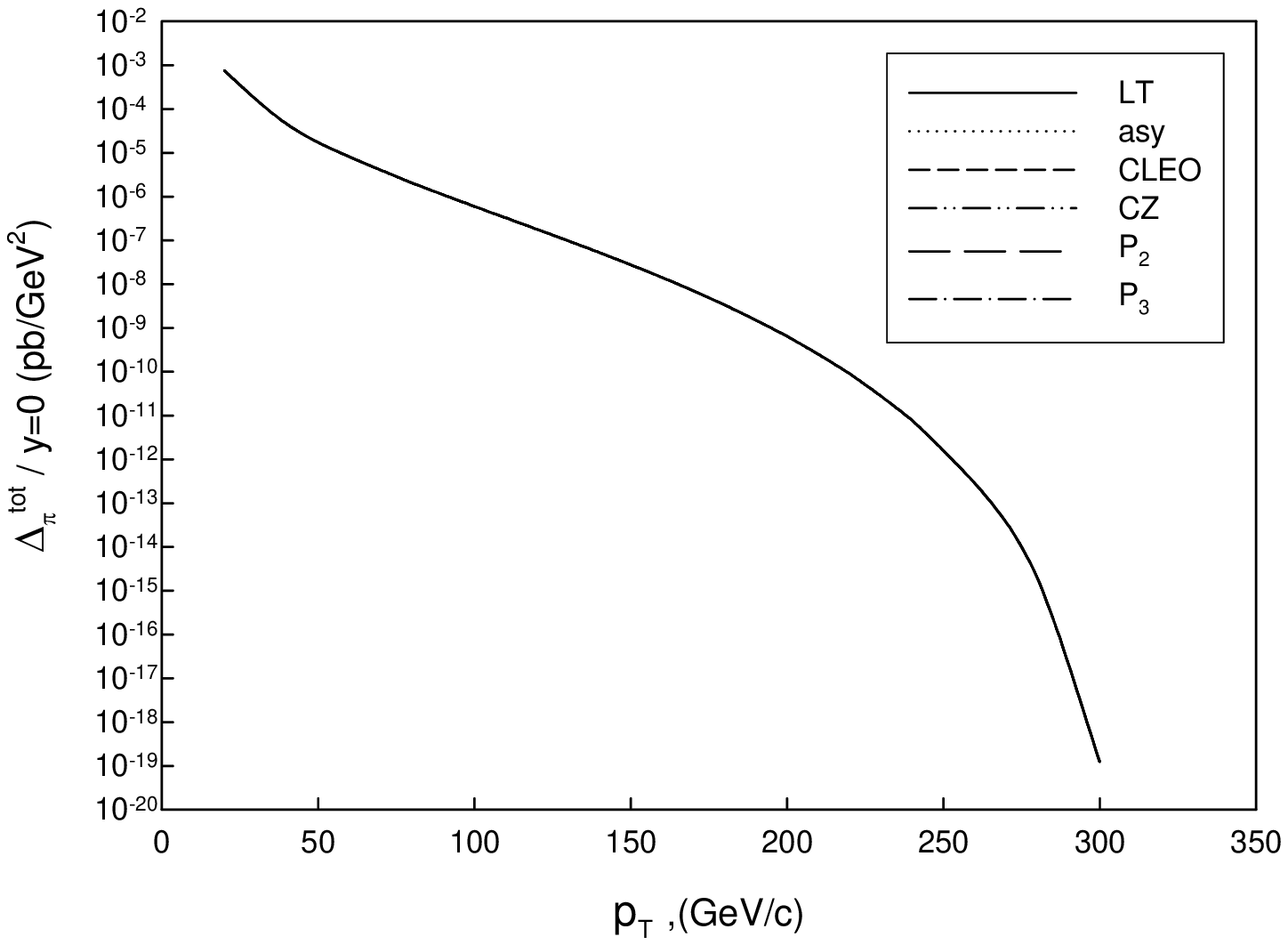}} \vskip-19.95cm
\caption{The sum of the difference of the leading and the high twist
cross sections,
$\Delta_{\pi}^{tot}$=$\Delta_{\pi}^{LT}$+$\Delta_{\pi}^{HT}$, as a
function of the  pion transverse momentum, $p_T$, at the c.m. energy
$\sqrt s=630GeV$.} \label{Fig15}
%\end{figure}
%\begin{figure}[h]
\epsfxsize 21.5cm \centerline{\epsfbox{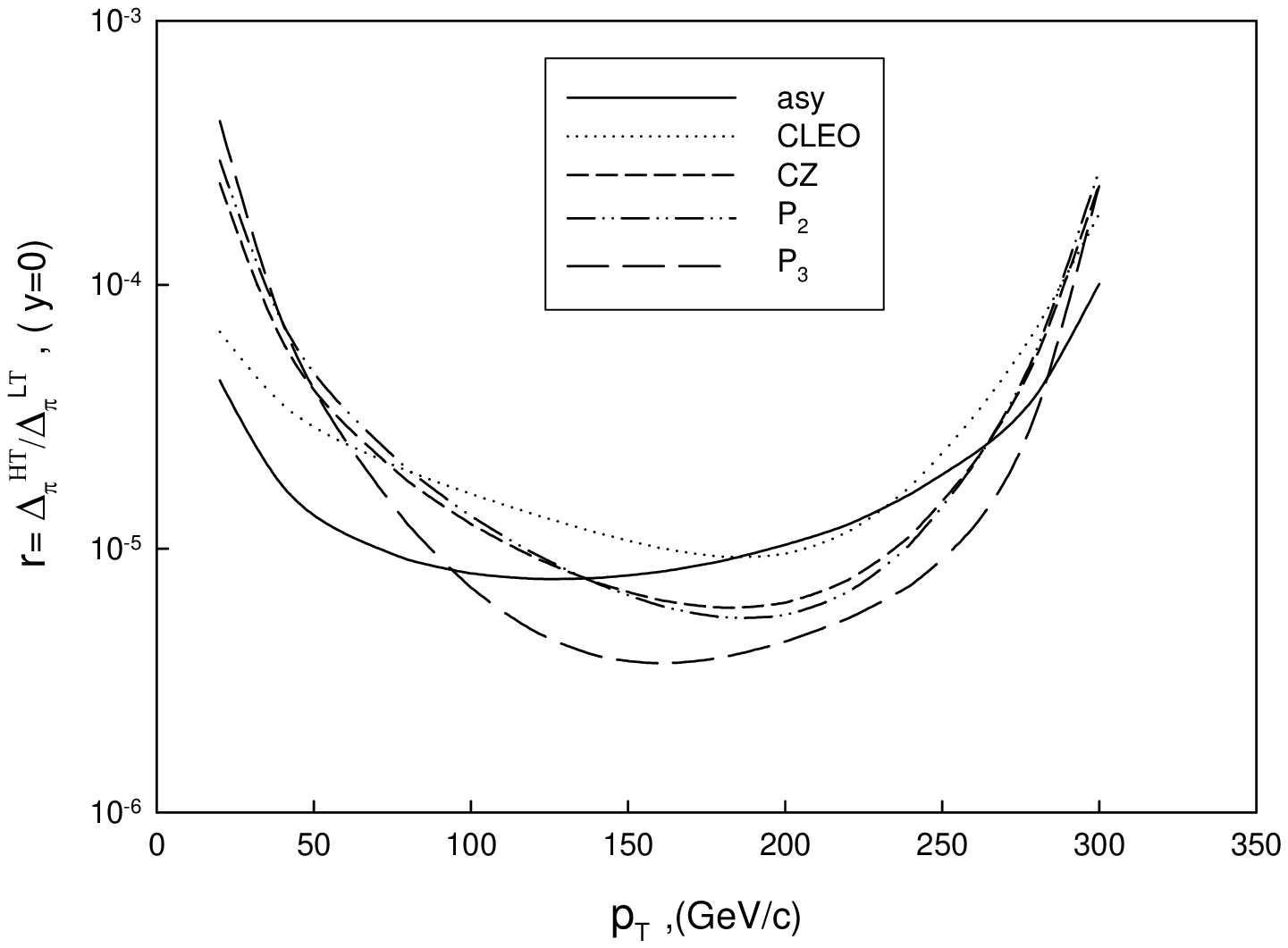}} \vskip-19.95cm
\caption{The ratio $r={\Delta_{\pi}^{HT}}/{\Delta_{\pi}^{LT}}$,
where the difference of the  leading and the high twist
contributions are calculated for the pion rapidity, $y=0$, at the
c.m. energy $\sqrt s= 630GeV$, as a function of the pion transverse
momentum $p_T$.} \label{Fig16}
\end{figure}

\newpage

\begin{figure}[htb]
 \vskip-2.0cm\epsfxsize 21.5cm \centerline{\epsfbox{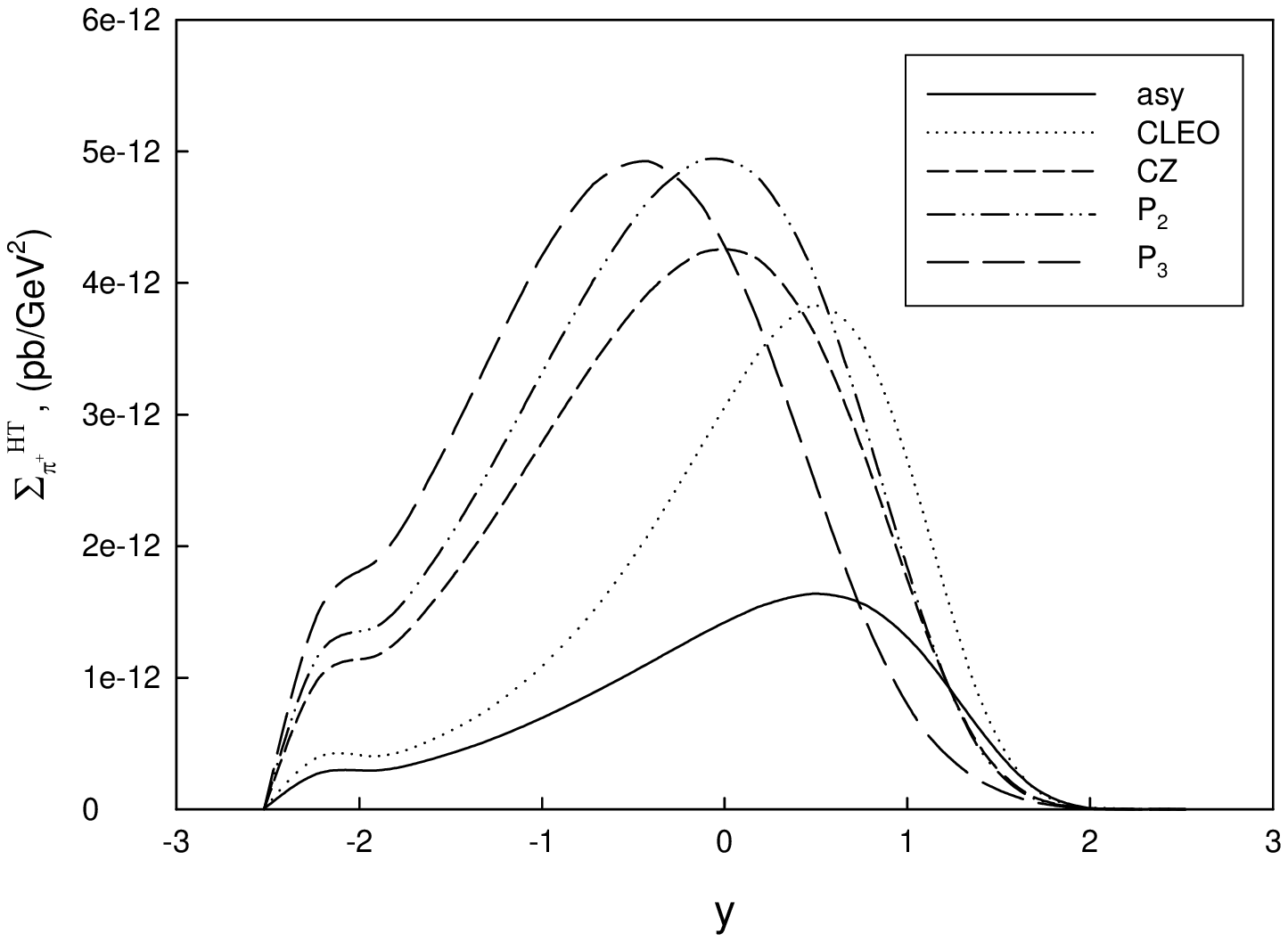}} \vskip-19.75cm
\caption{The high twist $\pi^{+}$ production cross sections as a
function of the $y$ rapidity of the pion at the transverse momentum
of the pion $p_T=50 GeV$, at c.m. energy $\sqrt s= 630GeV$.}
\label{Fig17}
%\end{figure}
%\begin{figure}[h]
\epsfxsize 21.5cm \centerline{\epsfbox{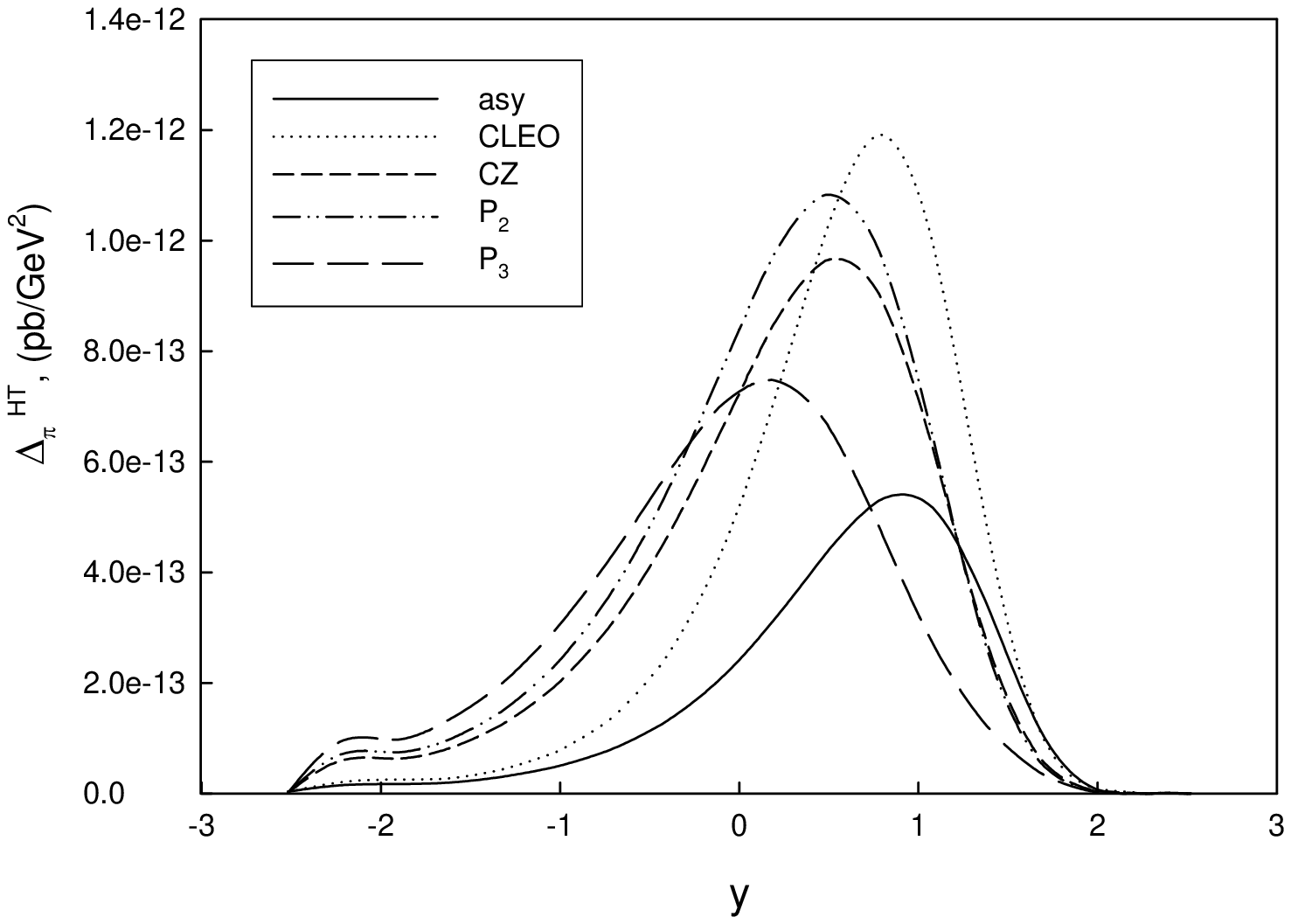}} \vskip-19.75cm
\caption{The difference of the high twist cross section,
$\Delta_{\pi}^{HT}$=$\Sigma_{\pi^{+}}^{HT}$-$\Sigma_{\pi^{-}}^{HT}$,
as a function of the $y$ rapidity of the pion at the  transverse
momentum of the pion $p_T=50 GeV$, at c.m. energy $\sqrt s=
630GeV$.} \label{Fig18}
\end{figure}

\end{document}